\documentclass[%
reprint,
superscriptaddress,
bibnotes,
longbibliography,
 amsmath,amssymb,
 prd,
]{revtex4-1}

\usepackage[colorlinks = true,
            linkcolor = blue,
            urlcolor  = blue,
            citecolor = blue,
            anchorcolor = blue]{hyperref}

\usepackage[caption=false]{subfig}

\usepackage{graphicx} 
\usepackage{dcolumn} 
\usepackage{bm} 

\usepackage{comment} 
\usepackage{todonotes} 

\usepackage{lineno} 

\usepackage{placeins} 

\usepackage{multirow} 

\usepackage{booktabs} 
\AtBeginDocument{
\heavyrulewidth=.08em
\lightrulewidth=.05em
\cmidrulewidth=.03em
\belowrulesep=.65ex
\belowbottomsep=0pt
\aboverulesep=.4ex
\abovetopsep=0pt
\cmidrulesep=\doublerulesep
\cmidrulekern=.5em
\defaultaddspace=.5em
}

\begin{document}

\title{Broadband solenoidal haloscope for terahertz axion detection}

\author{Jesse Liu}
\email{jesseliu@hep.phy.cam.ac.uk}
\affiliation{Cavendish Laboratory, University of Cambridge, Cambridge CB3 0HE, UK}
\affiliation{Department of Physics, University of Chicago, Chicago, IL 60637, USA}

\author{Kristin Dona}
\affiliation{Department of Physics, University of Chicago, Chicago, IL 60637, USA}

\author{Gabe Hoshino}
\affiliation{Department of Physics, University of Chicago, Chicago, IL 60637, USA}
\affiliation{Fermi National Accelerator Laboratory, Batavia, IL 60510, USA}

\author{Stefan Knirck}
\email{knirck@fnal.gov}
\affiliation{Fermi National Accelerator Laboratory, Batavia, IL 60510, USA}

\author{Noah Kurinsky}
\email{kurinsky@slac.stanford.edu}
\affiliation{SLAC National Accelerator Laboratory, Menlo Park, CA 94025, USA}
\affiliation{Fermi National Accelerator Laboratory, Batavia, IL 60510, USA}
\affiliation{Kavli Institute for Cosmological Physics, University of Chicago, Chicago, IL 60637, USA}

\author{Matthew Malaker}
\affiliation{Fermi National Accelerator Laboratory, Batavia, IL 60510, USA}

\author{David W. Miller}
\email{david.w.miller@uchicago.edu}
\affiliation{Department of Physics, University of Chicago, Chicago, IL 60637, USA}
\affiliation{Enrico Fermi Institute, University of Chicago, Chicago, IL 60637, USA}

\author{Andrew Sonnenschein}
\email{sonnensn@fnal.gov }
\affiliation{Fermi National Accelerator Laboratory, Batavia, IL 60510, USA}

\author{Mohamed H. Awida}
\affiliation{Fermi National Accelerator Laboratory, Batavia, IL 60510, USA}

\author{Peter S. Barry}
\affiliation{Argonne National Laboratory, Lemont, IL 60439, USA}
\affiliation{Kavli Institute for Cosmological Physics, University of Chicago, Chicago, IL 60637, USA}

\author{Karl K. Berggren}
\affiliation{Massachusetts Institute of Technology, Cambridge, MA 02139, USA}

\author{Daniel Bowring}
\affiliation{Fermi National Accelerator Laboratory, Batavia, IL 60510, USA}

\author{Gianpaolo Carosi}
\affiliation{Lawrence Livermore National Laboratory, Livermore, CA 94551, USA}

\author{Clarence Chang}
\affiliation{Argonne National Laboratory, Lemont, IL 60439, USA}
\affiliation{Kavli Institute for Cosmological Physics, University of Chicago, Chicago, IL 60637, USA}

\author{Aaron Chou}
\affiliation{Fermi National Accelerator Laboratory, Batavia, IL 60510, USA}

\author{Rakshya Khatiwada}
\affiliation{Fermi National Accelerator Laboratory, Batavia, IL 60510, USA}
\affiliation{Department of Physics, Illinois Institute of Technology, Chicago, IL 60616, USA}

\author{Samantha Lewis}
\affiliation{Fermi National Accelerator Laboratory, Batavia, IL 60510, USA}

\author{Juliang Li}
\affiliation{Argonne National Laboratory, Lemont, IL 60439, USA}

\author{Sae Woo Nam}
\affiliation{National Institute of Standards and Technology, Boulder, CO 80305, USA}

\author{Omid Noroozian}
\affiliation{NASA Goddard Space Flight Center, Grenbelt, MD 20771, USA}

\author{Tony X. Zhou}
\affiliation{Massachusetts Institute of Technology, Cambridge, MA 02139, USA}

\collaboration{BREAD Collaboration}

\begin{abstract}
We introduce the Broadband Reflector Experiment for Axion Detection (BREAD) conceptual design and science program. 
This haloscope plans to search for bosonic dark matter across the $[10^{-3},1]$\,eV ($[0.24, 240]$\,THz) mass range. 
BREAD proposes a cylindrical metal barrel to convert dark matter into photons, which a novel parabolic reflector design focuses onto a photosensor.
This unique geometry enables enclosure in standard cryostats and high-field solenoids, overcoming limitations of current dish antennas.
A pilot 0.7\,m$^{2}$ barrel experiment planned at Fermilab is projected to surpass existing dark photon coupling constraints by over a decade with one-day runtime.
Axion sensitivity requires $<10^{-20}$\,W/$\sqrt{\rm Hz}$ sensor noise equivalent power with a 10\,T solenoid and $10$\,m$^{2}$ barrel. 
We project BREAD sensitivity for various sensor technologies and discuss future prospects.
\end{abstract}

\maketitle
\section{Introduction}

Astrophysical evidence for dark matter (DM) is unambiguous~\cite{Rubin:1970zza,Tyson:1998vp,Tegmark:2003ud,Clowe:2006eq,Akrami:2018vks,Bertone:2004pz}, but its particle properties remain enigmatic.
Recent efforts are expanding bosonic DM searches for $m_\text{DM} \lesssim 1$\,eV masses~\footnote{Natural units are used, where the (reduced) Planck's constant $\hbar$, speed of light $c$, vacuum permittivity $\varepsilon_0$ and permeability $\mu_0$ are set to unity $\hbar = c = \varepsilon_0 = \mu_0 = 1$, as conventional in particle physics.} predicted by many extensions of the Standard Model (SM)~\cite{Arvanitaki:2009fg,Jaeckel:2010ni,Essig:2013lka,Baker:2013zta,Battaglieri:2017aum,Ahmed:2018oog,Irastorza:2018dyq}, complementing higher-mass searches~\cite{Akerib:2016vxi,Tan:2016zwf,Aprile:2018dbl,Agnese:2018col,DAMIC:2020cut,Izaguirre:2015yja,Boveia:2018yeb,Aad:2019qnd,Aaboud:2019rtt,ATLAS:2021kxv}.
Notably, the unobserved neutron electric dipole moment~\cite{Harris:1999jx,Baker:2006ts,Pendlebury:2015lrz,Abel:2020gbr} motivates the quantum chromodynamics (QCD) axion $a$ predicted by the Peccei-Quinn solution of the strong charge-parity problem~\cite{Peccei:1977hh,Wilczek:1977pj,Weinberg:1977ma}. 
Dark photons $A'$ are also sought-after candidates arising in string theory scenarios~\cite{Holdom:1985ag,Dienes:1996zr,Pospelov:2008zw,Goodsell:2009xc}. 
These states have compelling early-Universe production mechanisms and their field oscillations with frequency $\nu = m_\text{DM}/2\pi$ exhibit DM properties~\cite{Preskill:1982cy,Abbott:1982af,Dine:1982ah,Nelson:2011sf,Arias:2012az,Chadha-Day:2021szb}. 
Nonzero DM-photon couplings enable laboratory detection via electromagnetic (EM) effects. 

The most-sensitive detection strategy today is the radio-frequency resonant-cavity haloscope~\cite{Sikivie:1983ip,DePanfilis:1987dk,Wuensch:1989sa,Hagmann:1990tj}, where ADMX~\cite{Asztalos:2001tf,Asztalos:2001jk,Asztalos:2009yp,Wagner:2010mi,Du:2018uak,Braine2020,ADMX:2021nhd},
CAPP~\cite{Lee:2020cfj,Jeong:2020cwz,CAPP:2020utb}, HAYSTAC~\cite{Kenany:2016tta,Brubaker:2016ktl,Zhong:2018rsr,Backes:2020ajv} probe QCD axions within $[1.8, 24]\,\mu$eV masses. 
However, this strategy has long-standing obstacles from (i) narrow band tuning to unknown $m_\text{DM}$ and (ii) impractical high-mass scaling for $m_\text{DM} \gtrsim 40\,\mu$eV. 
Scan rates fall precipitously with photon frequency $R_\text{scan} \sim \nu^{-14/3}$~\cite{Backes:2020ajv} and the number of required resonators scales unfavorably with effective volume $\sim m_\text{DM}^3$. 
Proposed dielectric haloscopes could probe $[40, 400]\,\mu$eV~\cite{TheMADMAXWorkingGroup:2016hpc,Millar:2016cjp,Brun:2019lyf} and $[0.1,10]$\,eV~\cite{Baryakhtar:2018doz} masses, while topological insulators target $[0.7, 3.5]$~meV~\cite{Marsh:2018dlj}. 
Significant sensitivity gaps persist across $[10^{-4},1]$\,eV masses, favored by several theoretical scenarios~\cite{Graham:2015rva,Gorghetto:2020qws,Co:2020xlh}, motivating broadband approaches.

This Letter introduces the Broadband Reflector Experiment for Axion Detection (BREAD) conceptual design to search multiple decades of DM mass without tuning to $m_\text{DM}$.
BREAD proposes a novel experimental design that optimally realizes dish-antenna haloscopes~\cite{Horns:2012jf}.
Its hallmark is a cylindrical metal barrel for broadband DM-to-photon conversion with a coaxial parabolic reflector that focuses signal photons onto a sensor.
In contrast to existing dish antennas using spherical or flat surfaces~\cite{Suzuki:2015sza,Knirck:2018ojz,Tomita:2020usq,FUNKExperiment:2020ofv,BRASS:website}, our geometry is optimized for enclosure in standard cryostats and compact high-field solenoids.
This enhances signal to noise, ensures the emitting surface and magnetic field are parallel, and keeps costs practical.
We delineate the optical properties of this novel geometry with detailed ray tracing and numerical simulation. 
While photoconversion is broadband, photosensor performance governs final discovery reach.
We assess various state-of-the-art sensors and discuss the advances required in quantum sensing technology for next-generation devices to fulfill BREAD science goals with broad anticipated impact in astronomy and beyond~\cite{dhillon2017}. 

\begin{figure*}
    \centering
    \includegraphics[width=0.76\linewidth]{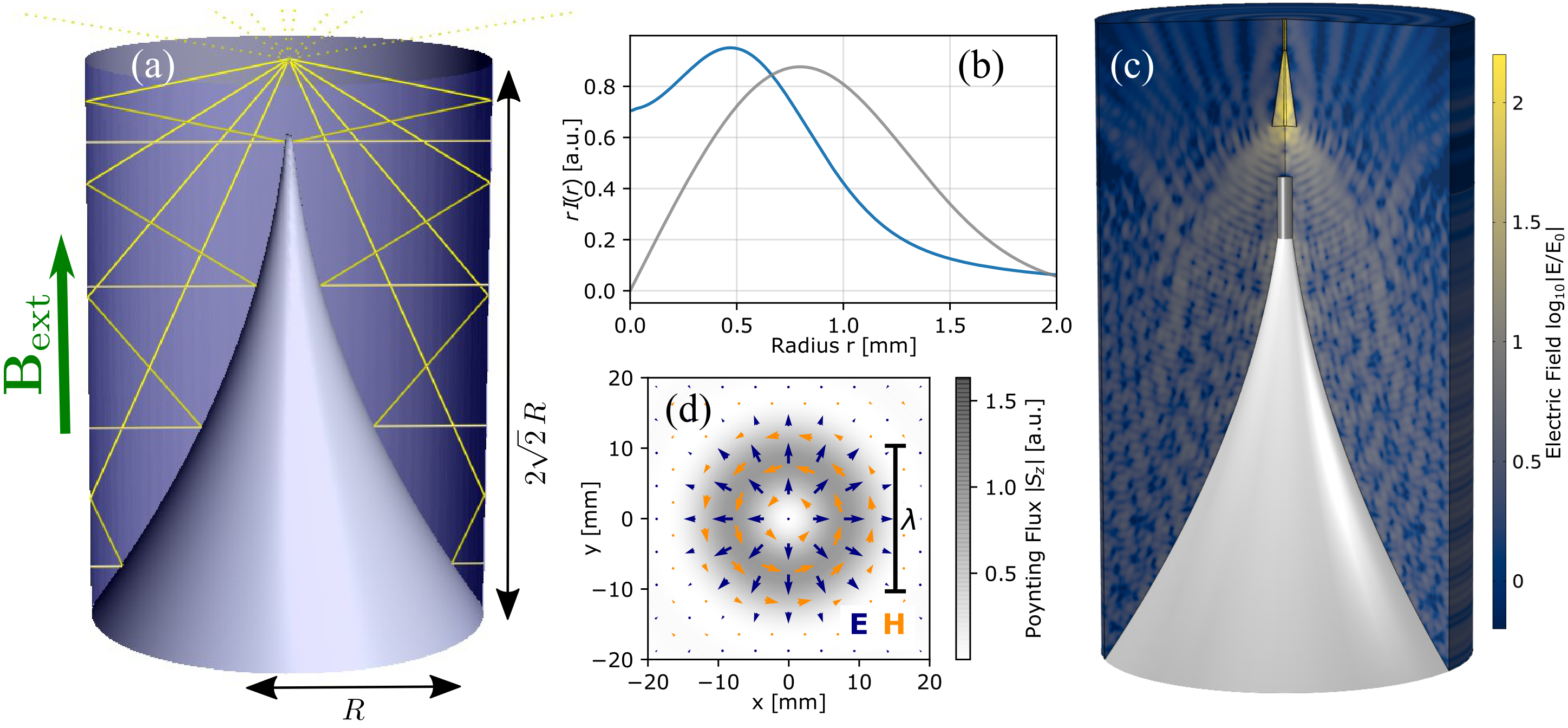}%
    \includegraphics[width=0.24\linewidth]{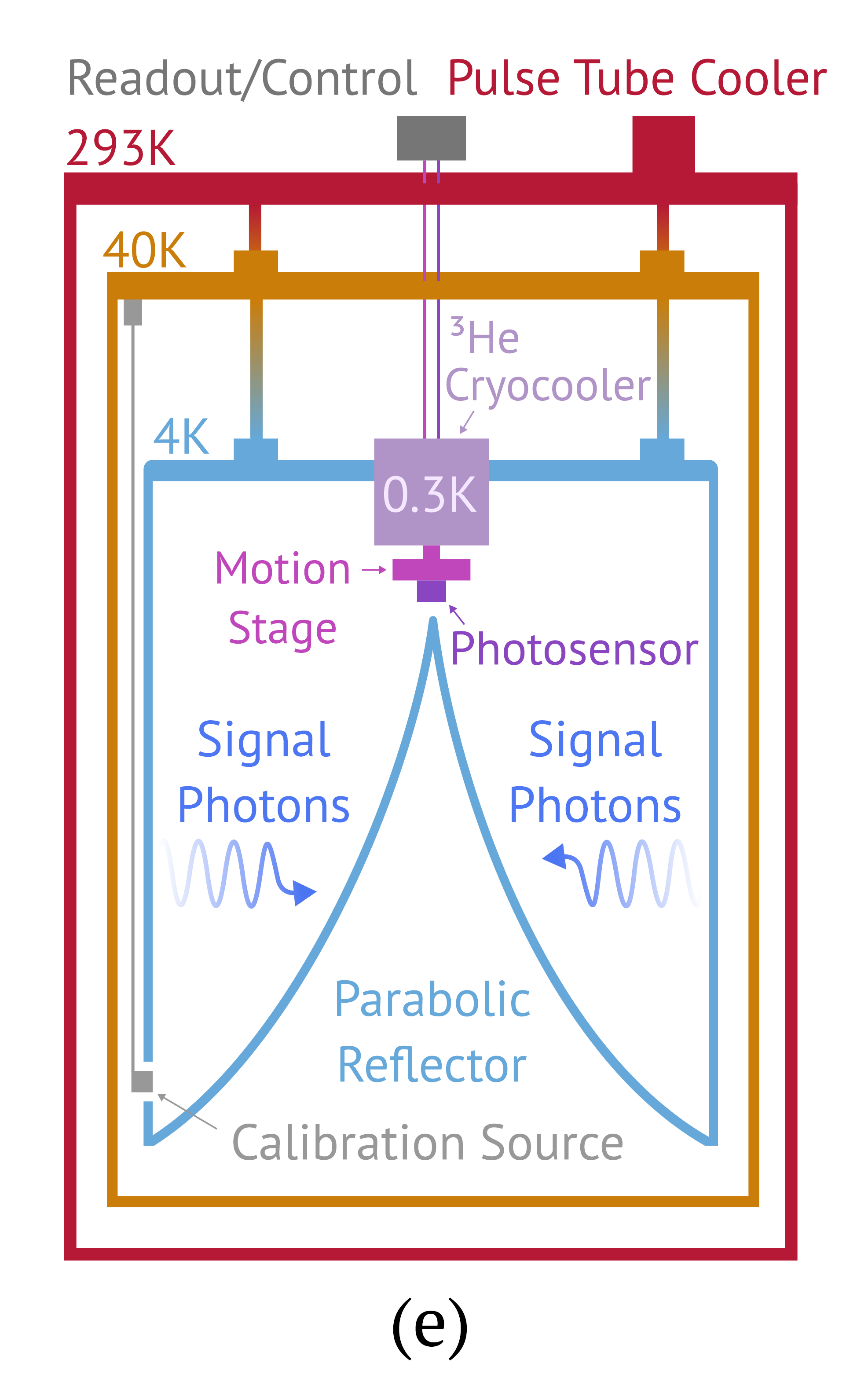}
    \caption{(a) BREAD reflector geometry: rays (yellow lines) emitted from the cylindrical barrel, which is parallel to an external magnetic field $\mathbf{B}_{\rm ext}$ from a surrounding solenoid (not shown) and focused at the vertex by a parabolic surface of revolution. 
    (b)~Radial intensity distribution $r I(r)$ expected from DM velocity effects in the $xy$ plane at the focal spot using ray tracing, for the BREAD geometry as in (a) with $R = 20\,{\rm cm}$ (blue) and for a conventional plane-parabolic mirror setup used in other experiments~\cite{Suzuki:2015sza,Knirck:2018ojz,Tomita:2020usq,FUNKExperiment:2020ofv,BRASS:website} with the same emitting surface area (gray). 
    (c) Full field simulation at around $15\,{\rm GHz}$ including a preliminary coaxial horn design. 
    (d) Electric (blue) and magnetic (orange) field distribution and time-averaged Poynting flux along the $z$ direction in the $xy$ plane at the focal spot. 
    (e) Schematic setup in cryostat for pilot dark photon searches.
    }
    \label{fig:simulation}
\end{figure*}

\vspace{-0.5cm}
\section{\label{sec:signal}Dark matter signal}
\vspace{-0.2cm}

Sub-eV-mass bosonic DM behave as classical fields, whose coherent oscillations generate the local halo energy density $\rho_\text{DM}$, which we assume to be $0.45$\,GeV\,cm$^{-3}$~\footnote{This value of $\rho_\text{DM}$ is typically adopted in axion haloscope literature~\cite{Du:2018uak}, but we note its significant uncertainties that range from $[0.2,0.6]$\,GeV\,cm$^{-3}$ from global methods to $[0.4,1.5]$\,GeV\,cm$^{-3}$ using recent astrometry data~\cite{Read:2014qva,ParticleDataGroup:2018ovx,Gaia:2018ydn,Buch:2018qdr}.}.
We consider scenarios where either axions or dark photons exclusively saturate the halo DM. 
The DM-photon interaction augments the Amp\`ere-Maxwell equation with an effective source current $\mathbf{J}_\text{DM}$~\cite{Jaeckel:2010ni} 
\begin{equation}
    \nabla \times \mathbf{B} -  \partial_t\mathbf{E} = \mathbf{J}_\text{DM}.
\end{equation}
A nonzero $\mathbf{J}_\text{DM}$ induces a small EM field that causes a discontinuity at the interface of media with different electric permittivity, such as a conducting dish in vacuum. 
To satisfy the $\mathbf{E}_{\parallel} = 0$ boundary condition parallel to the dish surface, a compensating EM wave with amplitude $|\mathbf{E}_0|$ must be emitted perpendicular to the surface. 
These waves transmit $P_\text{DM} = \frac{1}{2} |\mathbf{E}_0|^2 A_\text{dish}$ of power for dish area $A_\text{dish}$. 
For axions with $g_{a\gamma\gamma}$ coupling to photons, the current is $\mathbf{J}_a = g_{a\gamma\gamma} \sqrt{2\rho_\text{DM}} \mathbf{B}_\text{ext}^\parallel \cos(m_a t)$ given an external magnetic field $\mathbf{B}_\text{ext}^\parallel$ with nonzero component parallel to the plate, resulting in $P_a = \frac{1}{2} \rho_\text{DM} (B_\text{ext}^{\parallel} g_{a\gamma\gamma} / m_a)^2 A_\text{dish}$ emitted power~\cite{Horns:2012jf}.
QCD axion models~\cite{Dine:1981rt,Zhitnitsky:1980tq,Kim:1979if,Shifman:1979if,GrillidiCortona:2015jxo} relate $g_{a\gamma\gamma}$ to the mass by $g_{a\gamma\gamma} \sim 10^{-13} (m_a/\text{meV})$\,GeV$^{-1}$, giving $m_a$-independent power.
For dark photons with $A'$-SM kinetic mixing $\kappa$ and polarization $\mathbf{\hat{n}}$, the current is $\mathbf{J}_{A'} = \kappa m_{A'} \sqrt{2\rho_\text{DM}} \mathbf{\hat{n}} \cos(m_{A'} t)$, yielding $P_{A'} = \frac{1}{2} \rho_\text{DM} \kappa^2 A_\text{dish} \alpha_\text{pol}^2$ power. 
The factor $\alpha_\text{pol} = \sqrt{2/3}$ averages over $A'$ polarizations~\cite{Horns:2012jf}. 
$P_{A'}$ is $m_{A'}$-independent and persists even when $\mathbf{B}_\text{ext} = \mathbf{0}$.
Signal emission occurs independent of frequency in principle, allowing searches across several mass decades in single runs. 

Practically, DM-detection sensitivity also depends on the signal emission-to-detection efficiency $\epsilon_s$, photosensor noise equivalent power (NEP), and runtime $\Delta t$.
NEP is defined as the incident signal power required to achieve unit signal-to-noise ratio (SNR) in a one Hertz bandwidth.
We estimate sensitivity to $g_{a\gamma\gamma}$ and $\kappa$ (squared) as the SNR exceeding five
$\text{SNR} = (\epsilon_s P_\text{DM} \sqrt{\Delta t} )/\text{NEP} > 5$, where we assume sensors have sufficiently fast readout bandwidth $\mathcal{O}(100\,{\rm kHz})$: 
\begin{linenomath}
\begin{align}
    \left\{\!\!
    \begin{array}{c}
        \left(\frac{g_{a\gamma\gamma}}{10^{-11}}\right)^2   \\
         \left(\frac{\kappa}{10^{-14}}\right)^2  
    \end{array} 
    \!\!\right\}
    &= \left\{\!\!
    \begin{array}{c}
        \frac{1.9}{\text{GeV}^2}\!
    \left(\frac{m_a}{\text{meV}}
    \frac{10\,\text{T}}{B_\text{ext}}\right)^{2}  \\
       7.6\frac{2/3}{\alpha_\text{pol}^2}  
    \end{array} 
    \!\!\right\}
    \frac{10\,\text{m}^2}{A_\text{dish}}
    \left(\frac{\text{hour}}{\Delta t}\right)^{1/2}
    \nonumber\\
    & \!\!\!\!\!\! \times \!
    \frac{\text{SNR}}{5}
    \frac{0.5}{\epsilon_s} 
    \frac{\text{NEP}}{10^{-21}\,\text{W}/\sqrt{\text{Hz}}}
    \frac{0.45\,\text{GeV/cm}^3}{\rho_\text{DM}}.
    \label{eq:axion_dp_sensitivity}
\end{align}
\end{linenomath}
At high masses, shot noise is relevant due to insufficient signal photons $N_\text{signal} = (\epsilon_s P_\text{DM} \Delta t) / m_\text{DM} < 5$. 
For the nominal $A_\text{dish}=10$\,m$^2$, $B_\text{ext} = 10$\,T configuration, QCD axions induce a few 1\,eV photons week$^{-1}$ so month-long runtimes render shot noise subdominant for $m_\text{DM} \lesssim 1$\,eV.

In photon-counting regimes, sensors with dark count rate DCR detect photons emitted at rate $R_\text{DM} = P_\text{DM} / m_\text{DM}$. 
We use the counting-statistics significance $Z = N_\text{signal}/\sqrt{N_\text{noise}} =  (\epsilon_s R_\text{DM} \Delta t )/\sqrt{\text{DCR}\Delta t} > 5$ to estimate sensitivity in the background-limited regime. 
In the background-free photon-counting limit, the coupling sensitivity scales faster $g_{a\gamma\gamma}^{\rm sens} \propto (\Delta t)^{-1/2}$. 
With nominal photoconversion rates down to 1 photon per day, scaling as $R_{\gamma} \approx 10^{-5}(1\,\mathrm{eV}/m_{a})\,\mathrm{Hz}$, the photosensors considered are background limited. 
We thus constrain our projections to this scenario, where appendix~\ref{sec:signalRate} discusses requirements of background-free experiments.

\section{Coaxial haloscope design}

BREAD proposes a cylindrical barrel as the emitting surface and a novel reflector geometry comprising a coaxial parabolic surface of rotation around its tangent.
This focuses the emitted radiation to a photosensor located on-axis at the parabola's vertex as shown in Fig.~\ref{fig:simulation}\,(a). 
DM-to-photon conversion also occurs at the parabolic surface but is not focused on the vertex.
For a barrel with radius $R$ and length $L = 2\sqrt{2}R$, the effective emitting area is $A_\text{dish} = 2\pi R \, L$.
This aspect ratio suits enclosure in conventional high-field solenoid magnets and ensures $\mathbf{B}_\text{ext}$ is parallel to the emitting surface.
Such magnets are widely used in basic or applied applications with fields reaching $10\,{\rm T}$ or higher~\cite{BUDINGER2018509,Bird:2020nos}.

While photoconversion occurs regardless of $m_\text{DM}$, sensitivity is limited at high (low) masses by focusing (diffraction) effects.
Both effects broaden the focal spot and reduce the geometric signal efficiency due to finite photosensor size.
In the high-mass limit $\lambda_{\rm dB} \ll R$, DM-to-photon conversion occurs incoherently as the DM de Broglie wavelength $\lambda_{\rm dB}$ is smaller than the radius of the barrel $R \sim 1$\,m.
Here, the DM-halo velocity $v\simeq 10^{-3}$ smears out the focal spot size~\cite{Jaeckel:2013sqa,Jaeckel:2015kea,Jaeckel:2017sjb} on length scales larger than the signal-photon wavelength $\lambda_{\rm sig}$, rendering diffraction effects negligible.
The blue line in Fig.~\ref{fig:simulation}\,(b) shows the expected intensity distribution at the focal spot for the most conservative case where the DM wind points along the least favorable direction. 
The gray line refers to a planar conversion surface of the same area comparable to other dish-antenna experiments~\cite{Suzuki:2015sza,Knirck:2018ojz,Tomita:2020usq,FUNKExperiment:2020ofv,BRASS:website} with an on-axis parabolic mirror at $1\,{\rm m}$ distance. 
Since rays impinge the focal plane from a larger solid angle,
BREAD achieves improved focusing.

In the opposite low-mass limit $\lambda_{\rm dB} \gg R$,
such defocusing effects are negligible and the signal can be detected coherently.
Figure~\ref{fig:simulation}\,(c) shows the result of a COMSOL simulation at around $15\,{\rm GHz}$.
Here the full modified Maxwell wave equation is solved to verify that there are no spurious sources or resonances excited that may interfere destructively with the signal. 
Figure~\ref{fig:simulation}\,(d) shows the diffraction-limited electromagnetic fields at the focal plane. 
The electric field polarized along the radial direction can be picked up by coaxial horn antennas~\cite{Barros:2013coax,Bykov:2008coax}.
Receiver designs based on microwave and submillimeter astronomy projects could be considered for signal collection.
Proof-of-principle pilot experiments near both these limits are in preparation.
The radio-frequency pilot targeting 10s of GHz masses, called GigaBREAD, will be detailed in future work.

Figure~\ref{fig:simulation}\,(e) shows the proposed experimental design for the pilot $A'$ search at infrared (IR) frequencies, called InfraBREAD.
Cooling the conducting surfaces to 4\,K suppresses thermal noise and we identified a large cryostat at Fermilab built to test ADMX resonators, which will be available in 2022.
The barrel is constructed from aluminum with $A_\text{dish} = 0.7\,{\rm m}^2~(10\,{\rm m}^2)$ for the pilot (upgrade).
A 4\,K blackbody with $0.7\,{\rm m}^2$ area and unit emissivity emits $\sim 10^{-8}$\,W of power above 1\,THz (4\,meV).
Simulation shows that thermal radiation is evenly distributed across the focal plane and so suppressed by $A_\text{sens} / A_\text{dish} \sim 10^{-6}$ for active sensor area $A_\text{sens}$. 
For $A_\text{dish} = 0.7\,{\rm m}^2$, $A_\text{sens} = 0.5\times 0.5$\,mm$^2$ yields 50\% (25\%) signal efficiency $\epsilon_s$ for optimistic (pessimistic) DM-wind alignment; see appendix~\ref{apndx:velocity_focusing} for further discussion.
For absolute alignment of the photosensor in the reflector, we propose a piezoelectric motion stage to fine-tune the sensor position at the focus. 
Off focus, the signal is not enhanced. 
This enables \emph{in situ} noise measurements by moving the single photosensor off axis or installing a second off-axis photosensor. 
A monochromatic laser or bandpass-filtered blackbody source can inject photons via a small hole in the barrel for absolute calibration of the reflector-photosensor setup. 
A room-temperature spectrometer at UChicago is available to characterize sources and filters~\cite{Dona:2021iwt}.

Various upgrades and optimizations could be implemented to improve sensitivity of the proposed experimental concept.
A small secondary mirror near the focal point could guide the signal toward a low-field region where, e.g.\ a chopper and/or spectrometer could be installed.
The optics may optimize the radiation polarization and incident angle on the photosensors.
A detector array or photon imager~\cite{Zhao2017SinglephotonIB} could also provide spatial resolution to correlate any observed signals with the astrophysical DM distribution.
Specifically, the focal point and signal undergo diurnal and annual modulations due to the rotating DM velocity vector in the lab frame~\cite{Knirck:2018knd} and possibly $A'$ polarization~\cite{Caputo:2021eaa}.
The total $A_{\rm dish}$ could be increased within the same available volume
by combining the signals from an array of smaller BREAD-like barrels, but this increases complexity significantly.
Cosmic-ray muons are a suggested noise source for photon counters~\cite{snspd:snowmass2021loi}; \emph{in situ} vetoes at the sensor or barrel exterior, and/or underground operation are mitigation strategies.
Studying these options is deferred to future work.

\section{Photosensor technologies}

The expected DM signal rates and optical geometry imply stringent photosensor requirements: broad spectral response $\Delta E / E > 1$, ultralow noise NEP $< 10^{-20}\,{\rm W}\,{\rm Hz}^{-1/2}$ or DCR $< 10^{-3}$\,Hz, and millimeter-size active area.
Bolometers are promising because they directly measure absorbed photon power, only the absorbing material or structures limit spectral response, and are established technology with diverse applications~\cite{Richards1994,pirro2017}.
They comprise a thermally isolated absorbing element with low heat capacity, sensitive thermometer and weak link to a cold thermal bath.
They can measure photon energies from $10^{-5}\,{\rm eV}$~\cite{Kokkoniemi2020} to $10^6\,{\rm eV}$.  
Bolometers are typically insensitive to static (DC) radiation due to instrumental low-frequency ($1/f$) noise, so input signals must be time-modulated with, e.g.\ a chopper that shifts the signal to a frequency where $1/f$ noise is subdominant and SNR increases $\propto (\Delta t)^{1/2}$.

Photon-counting devices (photocounters) are potentially more sensitive than total-power bolometry, since simple signal-processing techniques, e.g.\ thresholding and pulse fitting, can suppress noise.
Small devices achieve nearly background-free single-photon counting at thresholds $\gtrsim 1$\,eV.
For lower energies, numerous devices exploit athermal breaking of Cooper pairs, including kinetic inductance detectors (KID), superconducting nanowire single photon detectors (SNSPD), and quantum capacitance detectors (QCDet)~\footnote{QCDet representing quantum capacitance detector avoids ambiguity with QCD denoting quantum chromodynamics.}.

\begin{table}[]
\caption{\label{tab:photosensors} Illustrative photosensor performance: spectral energy $E$, operating temperature $T_\text{op}$, active area $A_\text{sens}$.
Bolometers (photocounters) report noise equivalent power NEP (dark count rate DCR). 
}
\vspace{0.2cm}
\begin{tabular}{@{}lccccc@{}}
\toprule
Photosensor                                  & $\dfrac{E}{\text{meV}}$ & $\dfrac{T_\text{op}}{\text{K}}$ & $\dfrac{\text{NEP}}{\text{W}/\sqrt{\text{Hz}}}$   & $\dfrac{A_\text{sens}}{\text{mm}^2}$ \\ \midrule
\textsc{Gentec}~\cite{Gentec}                & [0.4, 120]   & 293    & $1\cdot10^{-8}$     & $\pi 2.5^2$ \\
\textsc{IR Labs}~\cite{IRLabs}               & [0.24, 248]  & 1.6    & $5\cdot 10^{-14}$   & $1.5^2$   \\
KID/TES~\cite{ridder2016,Baselmans2017}      & [0.2, 125]   & 0.3     & $2\cdot 10^{-19}$   & $0.2^2$              \\\midrule
QCDet~\cite{Echternach:2018,Echternach2021}  & [2, 125]     & 0.015  & $\frac{{\rm DCR}}{{\rm Hz}} = 4$        & $0.06^2$\\[0.1cm]
SNSPD~\cite{Hochberg:2019cyy,Verma:2020gso}  & [124, 830]   & 0.3    & $\frac{{\rm DCR}}{{\rm Hz}} = 10^{-4}$  & $0.4^2$               \\
\bottomrule
\end{tabular}
\end{table}

We now discuss specific technologies motivating the values in Table~\ref{tab:photosensors} assumed for our projections.
We display typical $A_\text{sens}$, but later set $\epsilon_s = 50\%$ for simplicity assuming sensor development will enable scaling to required sizes.
Room-temperature (\textsc{Gentec} pyroelectric~\cite{Gentec}) and cryogenic (\textsc{IR Labs} semiconducting thermistor~\cite{IRLabs}) devices exemplify commercial performance.

Superconducting titanium-gold transition edge sensors (TES)~\cite{Irwin1995ApPhL,irwin2005transition,gerrits2016} report down to $2 \times 10^{-19}\,{\rm W}\,{\rm Hz}^{-1/2}$ NEP in arrays of $200 \times 200\,\mu{\rm m}^2$ pixels~\cite{ridder2016}.
TESs have broad spectral response, where a molybdenum-gold device reporting $4 \times 10^{-19}\,{\rm W}\,{\rm Hz}^{-1/2}$ NEP covers 1--4\,$\mu$m to 160--960\,$\mu$m (eV to meV)~\cite{Goldie2011}.
Elsewhere, small 10\,$\mu$m superconducting--normal-metal junction bolometers report $2\times 10^{-20}\,{\rm W}\,{\rm Hz}^{-1/2}$ NEP~\cite{Kokkoniemi2019}, 
which may be promising if active areas are scalable to millimeters~\cite{Zhang_2021}.

\begin{figure*}
    \centering
    \includegraphics[width=0.49\linewidth]{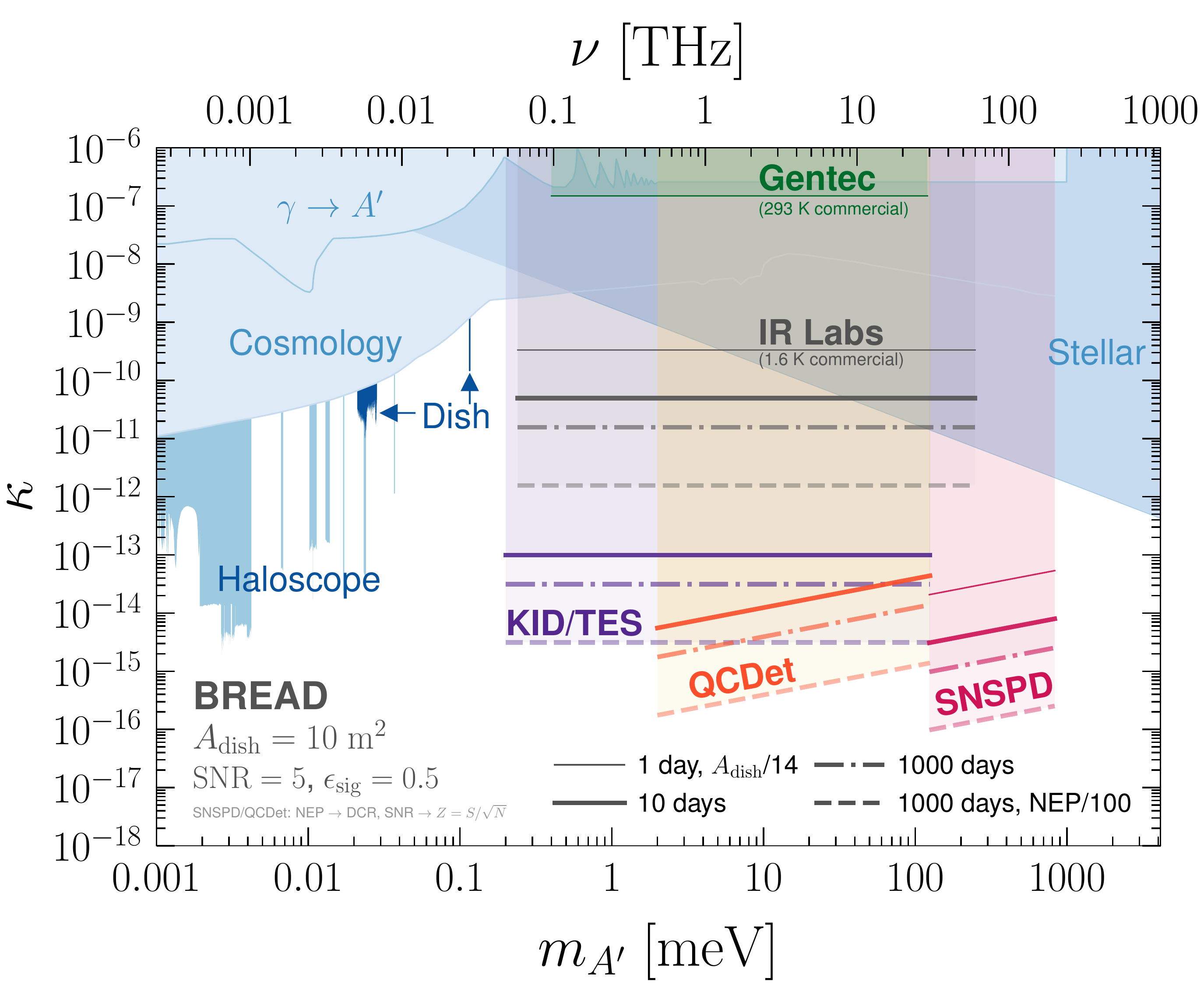}%
    \includegraphics[width=0.49\linewidth]{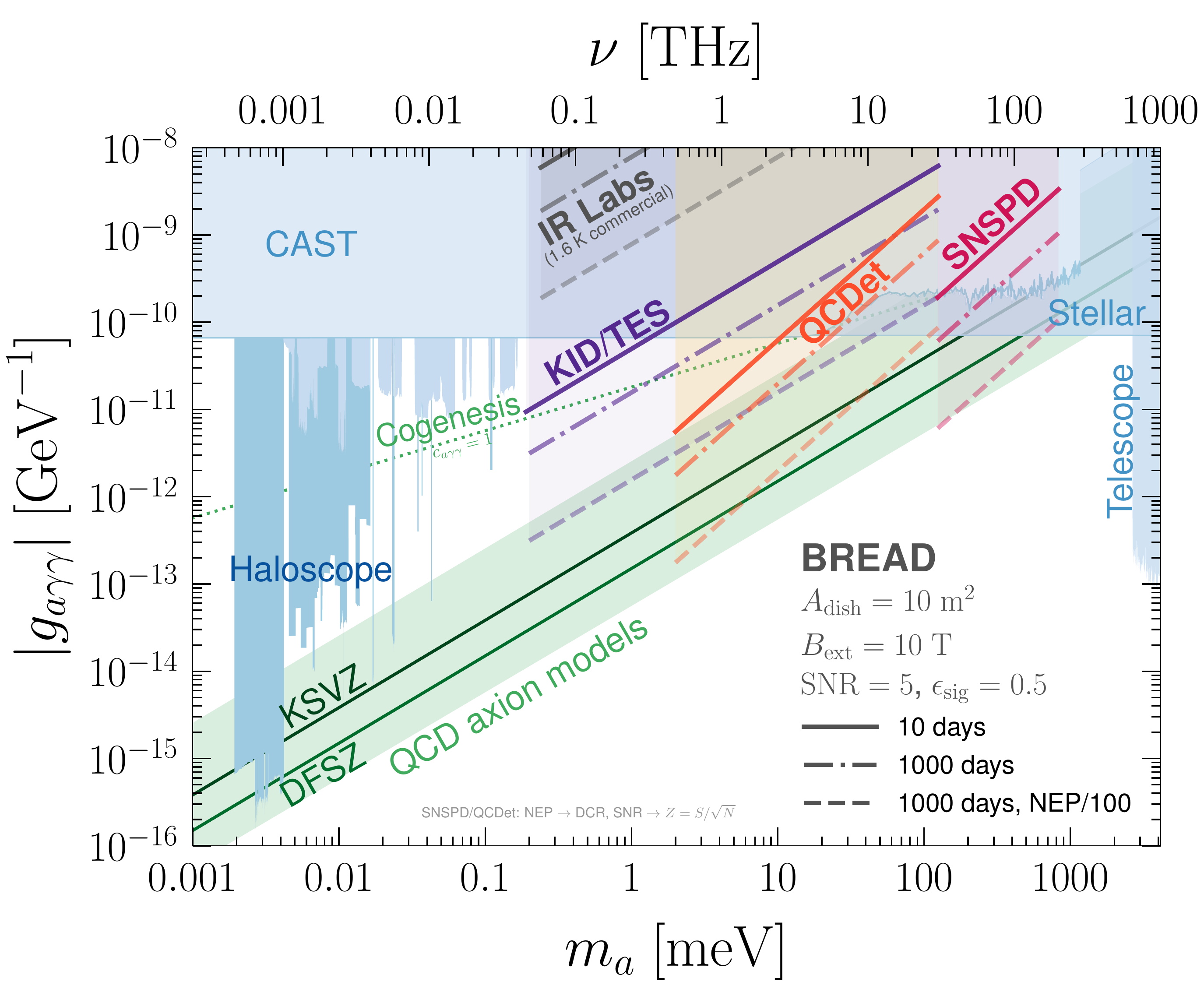}
    \caption{Projected BREAD sensitivity by sensor technology (bold labels, see Table~\ref{tab:photosensors}) for dark photons $A'$ (left) and axions $a$ (right).
    This assumes signal-to-noise ratio $\mathrm{SNR} = 5$ (significance $Z = 5$ for photocounters), signal efficiency $\epsilon_\text{sig} = 0.5$, and dish area $A_\text{dish} = 10$\,m$^{2}$. 
    Blue shading shows existing constraints from Ref.~\cite{Caputo:2021eaa}. 
    Benchmark axion predictions include QCD axion models~\cite{DiLuzio:2016sbl} (green band),  cogenesis~\cite{Co:2020xlh} (green dots), KSVZ~\cite{Kim:1979if,Shifman:1979if} and DFSZ~\cite{Dine:1981rt,Zhitnitsky:1980tq} (green lines). 
    Sensitivity scaling assumes background-limited operation where signal-power limits scale as $\sqrt{\Delta t}$ for runtime $\Delta t$ and linearly with improved NEP.
    }
    \label{fig:sensitivity_photosensors}
\end{figure*}

KIDs~\cite{Day2003,Leduc2010,doi:10.1146/annurev-conmatphys-020911-125022} are thin-film resonators, whose surface inductance is sensitive to Cooper-pair-breaking photons above the band gap $\Delta \simeq 0.2\,$meV.
Titanium-nitride KIDs are scalable to $50\times50$\,mm$^2$ kilopixel arrays with $3\times 10^{-19}\,{\rm W\,Hz}^{-1/2}$ NEP~\cite{Baselmans2017}, which are antenna coupled and optimized to [3.4, 12]\,meV~\cite{spica2012}. 
For cosmic microwave background (CMB) applications ($0.2\lesssim E \lesssim 2$\,meV), KIDs are limited by signal power rather than sensor noise at ${\rm NEP}\sim 10^{-17}\,{\rm W\,Hz}^{-1/2}$~\cite{abitbol2017cmb}, and therefore could have better performance in such frequencies than current sensors targeting CMB science. 
Given KID and TES devices report similar NEP in each application, we amalgamate their presentation in our projections for simplicity.
We extrapolate the $2\times 10^{-19}\,{\rm W\,Hz}^{-1/2}$ NEP~\cite{ridder2016} into the [0.2, 125]\,meV range where we expect KID/TES devices to operate bolometrically, but this will require experimental demonstration. 

QCDets~\cite{Shaw2009prb,bueno2010,Echternach2013} recently report $3 \times 10^{-21}\,{\rm W}\,{\rm Hz}^{-1/2}$ NEP at 1.5\,THz (6.2\,meV)~\cite{Echternach:2018}. 
These are scalable to 441 pixel arrays and simulation indicates 1-4$\times 10^{-20}\,{\rm W}\,{\rm Hz}^{-1/2}$ NEP for [2, 125]\,meV~\cite{Echternach2021}, driven by  e.g.\ Origins Space Telescope goals~\cite{Leisawitz2021}.
Such performance is promising, and for simplicity, we assume constant $3\times 10^{-21}\,{\rm W}\,{\rm Hz}^{-1/2}$ NEP in our projections. 
We convert this to ${\rm DCR} = 4\,{\rm Hz}$ using $\text{NEP} = (E/\eta)\sqrt{2\cdot\text{DCR}}$~\cite{Chen:18} for $E=6.2\,{\rm meV}$ and optical efficiency $\eta = 0.9$~\cite{Echternach:2018}. 

SNSPDs~\cite{Goltsman2001,Natarajan:2012bw,Chen:18} comprise sub-micron-width wires wound across thin-film substrates that count photons above an energy threshold.
Superconductivity is momentarily lost upon photon absorption, leading to a measurable voltage pulse.  
Such devices achieve $> 90\%$ efficiency~\cite{Marsili2013} and recently, a $400 \times 400$\,$\mu$m$^{2}$ tungsten-silicide device reports DCR $<10^{-4}$\,Hz for 0.8\,eV threshold~\cite{Hochberg:2019cyy}.
Using Fermilab refrigerators~\cite{Hernandez:2020hcj}, we are preparing to test similar SNSPDs fabricated at MIT. 
Recent advances important for BREAD include extending up to 10\,$\mu$m (0.12\,eV)~\cite{Verma:2020gso} and developing large $3.1\times 3.1$\,mm$^2$ single pixels~\cite{Wollman2021}.
Continued research to lower thresholds is motivated given axions with $m_a \lesssim 60\,{\rm meV}$ are disfavored by supernova constraints~\cite{Chang:2018rso}.

Photocounting is also possible using KIDs~\cite{Gao:2012rb,deVisser:2021kip} and TESs~\cite{MillerA2003,Lita:08,2012ApPhL.101e2601K}, with the benefit of per-photon energy resolution. 
With e.g.\ 10\% energy resolution determined by detector resolution, the monochromatic DM signal occupies one energy bin but noise can be spread across 10 bins, improving SNR after sufficient $\Delta t$ up to a trials factor.
Exploring this in BREAD requires more detailed resolution and noise models, which is deferred to future work. 

\section{Sensitivity and discussion}

We project BREAD sensitivity to dark photons in
Fig.~\ref{fig:sensitivity_photosensors}\,(left) using Eq.~\eqref{eq:axion_dp_sensitivity} assuming the spectral and noise benchmarks in Table~\ref{tab:photosensors}.
Existing constraints following Ref.~\cite{Caputo:2021eaa} (blue shading) include
stellar astrophysics~\cite{Redondo:2013lna,vinyoles2015new}, cosmology~\cite{Arias:2012az,McDermott:2019lch,Caputo:2020bdy}, 
and $\gamma\to A'$ conversion that includes laboratory probes~\cite{Betz:2013dza,Kroff:2020zhp}.
With just $1$\,day runtime assuming $A_\text{dish} = 0.7$\,m$^{2}$, the gray thin line shows the BREAD pilot could surpass existing $\kappa$ constraints by one decade around 1\,meV using the commercial \textsc{IR Labs} sensor.
The reach of BREAD is substantially broader compared with two existing dish antennas, SHUKET~\cite{Brun:2019kak} and Tokyo~\cite{Tomita:2020usq} (dark blue).
Importantly, BREAD probes higher masses than existing haloscopes ADMX~\cite{Asztalos:2001jk,Asztalos:2009yp},
CAPP~\cite{Lee:2020cfj,Jeong:2020cwz,CAPP:2020utb},
HAYSTAC~\cite{Zhong:2018rsr,Backes:2020ajv},
transmon qubit~\cite{Dixit:2020ymh}, and
WISPDMX~\cite{Nguyen:2019xuh}, whose results are recasted for $A'$ following Ref.~\cite{Caputo:2021eaa}.
Scaling to $A_\text{dish} = 10$\,m$^{2}$ and using KID/TES sensors could open two decades further $\kappa$ sensitivity, while SNSPDs could achieve three decades gain for $m_{A'}\gtrsim 0.1$\,eV.
Extending runtime to $\Delta t = 10^3$\,days enables $\kappa$ sensitivity to reach six (four) decades beyond existing constraints for $m_{A'} \sim 0.4~(200)$\,meV.

Axion sensitivity is illustrated in
Fig.~\ref{fig:sensitivity_photosensors}\,(right). 
Existing constraints~\cite{Caputo:2021eaa} additionally include the
CAST helioscope~\cite{Arik:2013nya,Anastassopoulos:2017ftl}, 
telescopes~\cite{Grin:2006aw,Regis:2020fhw}, 
neutron stars~\cite{Foster:2020pgt,Darling:2020uyo,Battye:2021yue}, 
alongside
ORGAN~\cite{McAllister:2017lkb}
QUAX~\cite{Alesini:2019ajt,Alesini:2020vny}
RADES~\cite{CAST:2021add} and
URF~\cite{DePanfilis:1987dk,Wuensch:1989sa,Hagmann:1990tj} haloscopes.
While challenging with commercial devices, 10\,day runtimes using KID/TES sensors with NEP $\sim 10^{-19}\,{\rm W}\,{\rm Hz}^{-1/2}$ could surpass CAST sensitivity for $m_a \lesssim 10$\,meV.
Longer runtimes could test cogenesis predictions for the $c_{a\gamma\gamma}=1$ benchmark~\cite{Co:2020xlh}.
Increasing $A_\text{dish}$ ($B_\text{ext}$) beyond 10\,m$^2$ (10\,T) is financially unfavorable, requiring custom cryostats and magnets.
Thus practically probing QCD axion models~\cite{DiLuzio:2016sbl} requires longer runtime and lower sensor noise.
Coupling sensitivity scales slowly with runtime $g_{a\gamma\gamma}^\text{sens} \sim (\Delta t)^{-1/4}$, i.e.\ halving $g_{a\gamma\gamma}^\text{sens}$ requires $16 \times$ longer runtimes. 
For $\Delta t = 10^3$\,days, reaching KSVZ~\cite{Kim:1979if,Shifman:1979if} (DFSZ~\cite{Dine:1981rt,Zhitnitsky:1980tq}) demands $1~(0.2)\times 10^{-22}\,{\rm W}\,{\rm Hz}^{-1/2}$ NEP. 
Achieving this NEP for wide spectral ranges is challenging and a key science driver for sensor development.
This may be attainable above 0.1\,meV for photocounters, e.g.\ SNSPDs, motivating dedicated measurements in preparation, and next-generation bolometers at lower masses given a recent TES-based device reports $8 \times 10^{-22}\,{\rm W}\,{\rm Hz}^{-1/2}$ electrical NEP~\cite{Nagler:2020mki}.
Maintaining signal efficiency when upgrading $A_\text{dish} = 0.7 \to 10$\,m$^{2}$ requires quadrupling the active sensor width.
Overcoming these challenges promises significant scientific payoff given the multidecade improvements in search coverage that has long eluded cavity haloscopes. 
Post discovery, the DM signal will always persist, enabling cross checks with resonant techniques and measurements to elucidate its particle physics and astrophysical properties~\cite{Knirck:2018ojz,Caputo:2021eaa}. 

In summary, we proposed BREAD to improve sub-eV-mass DM reach by several decades.
We introduced the novel coaxial design optimized for embedding in standard solenoids and cryostats, in contrast to existing dish antennas, then detailed numerical optics simulation and examined photosensor candidates.
Realizing BREAD into a cornerstone DM experiment will catalyze synergies across quantum technology and astroparticle physics. 

\subsection*{Acknowledgments}

We thank Ankur Agrawal, Israel Alatorre, Kate Azar, Lydia Beresford, Pierre Echternach, Juan Estrada, Amanda Farah, Casey Frantz, Mary Heintz, Chris Hill, Matthew Hollister, Reina Maruyama, Jan Offermann, Mark Oreglia, Jessica Schmidt, Sadie Seddon-Stettler, Danielle Speller, Emily Smith, Cecilia Tosciri, Liliana Valle, Joaquin Viera, and Steven Zoltowski for interesting and helpful discussions.
This work is funded in part by the Department of Energy through the program for Quantum Information Science Enabled Discovery (QuantISED) for High Energy Physics and the resources of the Fermi National Accelerator Laboratory (Fermilab), a U.S. Department of Energy, Office of Science, HEP User Facility. 
Fermilab is managed by Fermi Research Alliance, LLC (FRA), acting under Contract No. DE-AC02-07CH11359.
Work at Argonne National Laboratory is supported by the U.S. Department of Energy, Office of High Energy Physics, under contract DE-AC02-06CH11357.
We acknowledge support by the Kavli Institute for Cosmological Physics at the University of Chicago through grant NSF PHY-1125897 and an endowment from the Kavli Foundation and its founder Fred Kavli.
Work at Lawrence Livermore National Laboratory is supported under contract number DE-AC52-07NA27344; Release \#LLNL-JRNL-828670. 
The MIT co-authors acknowledge support from the Fermi Research Alliance, LLC (FRA) and the US Department of Energy (DOE) under contract No. DE-AC02-07CH11359.
We thank the Aspen Center for Physics, which is supported by National Science Foundation grant PHY-1607611, for hosting the Quantum Information Science for Fundamental Physics meeting (2020), and the University of Washington CENPA for the Axions Beyond Gen2 workshop (2021) supported by the Heising-Simons Foundation. 
JL acknowledges a University of Chicago fellowship supported by the Grainger Foundation, where this work began, and a Junior Research Fellowship at Trinity College, University of Cambridge.

\bibliography{refs.bib}


\appendix 
\clearpage
\FloatBarrier
\section*{Appendix}

This appendix first reviews the axion and dark photon signal rate and sensitivity in bolometric and photocounting regimes.
Next, we present additional discussion on the cryostat and magnet followed by dark matter velocity effects on the signal spot size. 
Then, we expand discussion on photosensor performance and noise sources before commenting further on bolometers.
Finally, we discuss a potential experimental staging for BREAD. 

\subsection{Signal rate}\label{sec:signalRate}
The interaction Lagrangian coupling SM photons to dark photons $A'$ (a spin 1 vector) and axions $a$ (a spin 0 pseudoscalar) is given respectively by~\cite{Jaeckel:2010ni,Horns:2012jf}
\begin{equation}
    \mathcal{L}_{A'} = -\frac{1}{4} \kappa F_{\mu\nu}'F^{\mu\nu},\quad
    \mathcal{L}_{a} = -\frac{1}{4} g_{a\gamma\gamma} a F_{\mu\nu}\tilde{F}^{\mu\nu},
    \label{eq:intLagrangian}
\end{equation}
where $\tilde{F}^{\mu\nu} = \epsilon^{\mu\nu\rho\sigma}F_{\rho\sigma}$ with $\epsilon^{\mu\nu\rho\sigma}$ being the totally antisymmetric tensor,  $g_{a\gamma\gamma}$ is the axion--photon coupling, 
$\kappa$ is the kinetic mixing parameter, 
and $F_{\mu\nu}^{(')} = \partial_{[\mu}A^{(')}_{\nu]}$ is the SM (dark) photon field strength.
Upon solving the Lagrangian equations of motion, the consequential modification to SM electrodynamics is an additional effective source current $\mathbf{J}_\text{DM}$ in the Amp\`ere--Maxwell equation~\cite{Sikivie:1983ip}
\begin{equation}
\nabla \times \mathbf{B} -  \partial_t\mathbf{E} = \mathbf{J}_\text{DM}.
\label{eq:axionAmpereMaxwell}
\end{equation}
The nonzero current $\mathbf{J}_\text{DM}$ from the DM--photon interaction Lagrangian of Eq.~\eqref{eq:intLagrangian} is given by 
\begin{equation}
\mathbf{J}_{A'} = \kappa m_{A'}^2\mathbf{A}'(t),\quad
\mathbf{J}_{a}  = g_{a\gamma\gamma}\mathbf{B}_\text{ext} m_a a(t),
\label{eq:axion_dp_current}
\end{equation}
where $\mathbf{A}'(t)$ and $a(t)$ are the dynamical dark photon and axion fields, respectively, and $\mathbf{B}_\text{ext}$ is the external magnetic field applied in the laboratory.
In general, one considers wave solutions of the form $\exp[\mathrm{i}(\mathbf{k}\cdot \mathbf{x} - \omega t)]$, where $\omega = m_\text{DM}$.
As the local DM halo has nonrelativistic velocity $v \simeq 10^{-3}$, the momentum wavevector approximately vanishes $\mathbf{k} \to 0$ and thus $\nabla \times \mathbf{B}$ in Eq.~\eqref{eq:axionAmpereMaxwell} induced by DM is negligible.
From the cosmology of bosonic condensate DM, the local halo energy density $\rho_\text{DM}$ is related to the DM mass and fields by
\begin{equation}
    \rho^{(A')}_\text{DM} = \frac{1}{2}m_{A'}^2 |\mathbf{A}'(t)|^2, \quad
    \rho^{(a)}_\text{DM}  = \frac{1}{2}m_{a}^2 a^2(t).
\end{equation}
Using this in Eq.~\eqref{eq:axion_dp_current} gives
the resulting source current for axions has the form $\mathbf{J}_a = g_{a\gamma\gamma} \sqrt{2\rho_\text{DM}} \mathbf{B}_\text{ext} \cos(m_a t)$. 
For dark photons, the source current is
$\mathbf{J}_{A'} = \kappa m_{A'} \sqrt{2\rho_\text{DM}} \mathbf{\hat{n}} \cos(m_{A'} t)$, where $\mathbf{\hat{n}}$ is the $A'$ polarization. 
The electric field induced by $\mathbf{J}_\text{DM}$ is then deduced from Eq.~\eqref{eq:axionAmpereMaxwell} and has magnitude
\begin{equation}
    |\mathbf{E}_{A'}| = \kappa\sqrt{2\rho_\text{DM}},\quad
    |\mathbf{E}_{a}|  = \left(g_{a\gamma\gamma}|\mathbf{B}_\text{ext}^{||}|/m_a\right)\sqrt{2\rho_\text{DM}},
    \label{eq:DMinducedEfield}
\end{equation}
where $\mathbf{B}_\text{ext}^{||}$ is the magnetic field component parallel to the emitting barrel surface. 
This nonzero electric field causes a discontinuity at the interface of a conducting dish in vacuum. 
To satisfy the $\mathbf{E}_{\parallel} = 0$ boundary condition parallel to the dish surface (from Faraday's law $\nabla \times \mathbf{E} + \partial_t \mathbf{B} = 0$, which remains unchanged), a compensating electromagnetic wave with amplitude $\mathbf{E}_0$ must be emitted perpendicular to the surface. 
The power per unit area $P/A_\text{dish}$ of the emitted waves is given by $P/A_\text{dish} = \frac{1}{2} |\mathbf{E}_0|^2$ and applying this to the axion-induced electric field of Eq.~\eqref{eq:DMinducedEfield} gives $P_a = \frac{1}{2} \rho_\text{DM} (B_\text{ext}^{\parallel} g_{a\gamma\gamma} / m_a)^2 A_\text{dish}$ for dish area $A_\text{dish}$.
Normalizing the power to experimental parameters, the power emitted for axion and dark photon DM is, respectively
\begin{linenomath}
\begin{align}
    \left\{\!
    \begin{array}{c}
        \dfrac{P_a   }{8.8 \cdot 10^{-23}\,\text{W}}  \\[0.3cm]
        \dfrac{P_{A'}}{2.2 \cdot 10^{-23}\,\text{W}}
    \end{array} 
    \!\right\}
    &= \left\{\!\!
    \begin{array}{c}
	    \left(\frac{g_{a\gamma\gamma}}{10^{-11}\,\text{GeV}^{-1}} 
	    \frac{\text{meV}}{m_a}\right)^2
        \left(\dfrac{B_\text{ext}}{10\,\text{T}}\right)^2 \\[0.3cm]
        \dfrac{\alpha_\text{pol}^2}{2/3} \left(\dfrac{\kappa}{10^{-14}}\right)^2
    \end{array} 
    \!\!\right\}
    \nonumber\\
    &  \quad \times 
    \frac{\rho_\text{DM}}{0.45\,\text{GeV/cm}^3}
	\frac{A_\text{dish}}{10\,\text{m}^2}.
    \label{eq:axion_dp_rate}
\end{align}
\end{linenomath}

\begin{figure}
    \centering
    \includegraphics[width=\linewidth]{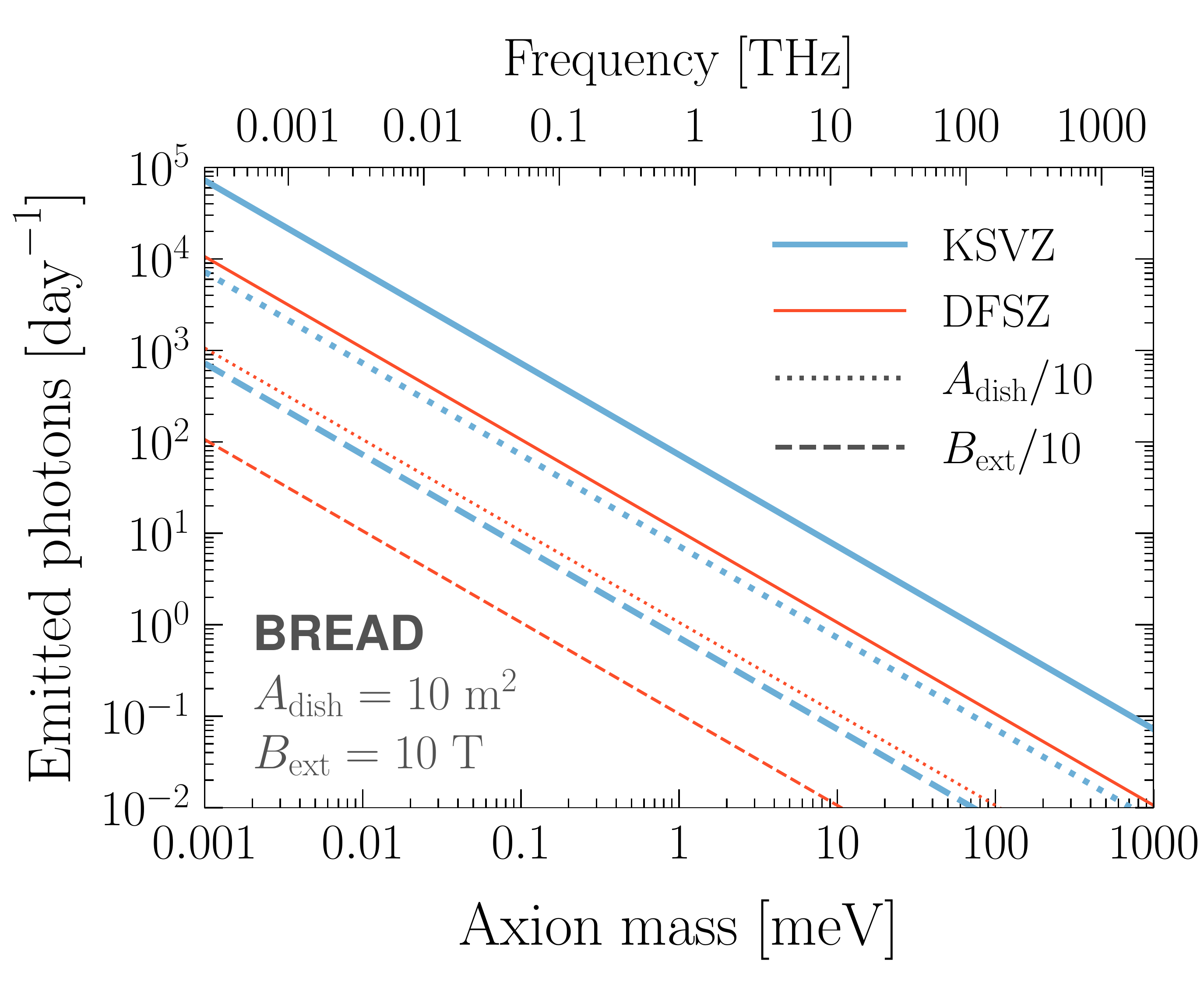}
    \caption{Photon emission rate of QCD axion benchmarks using Eq.~\eqref{eq:summary_qcd_axion_rate}. 
    Thick blue (thin orange) solid lines shows the KSVZ (DFSZ) models. 
    Dotted (dashed) variations shows the dish area (magnetic field) reduced by a factor of 10.}
    \label{fig:photon_rate_qcdaxion}
\end{figure}

\begin{figure*}
    \centering
    \includegraphics[width=0.33\linewidth]{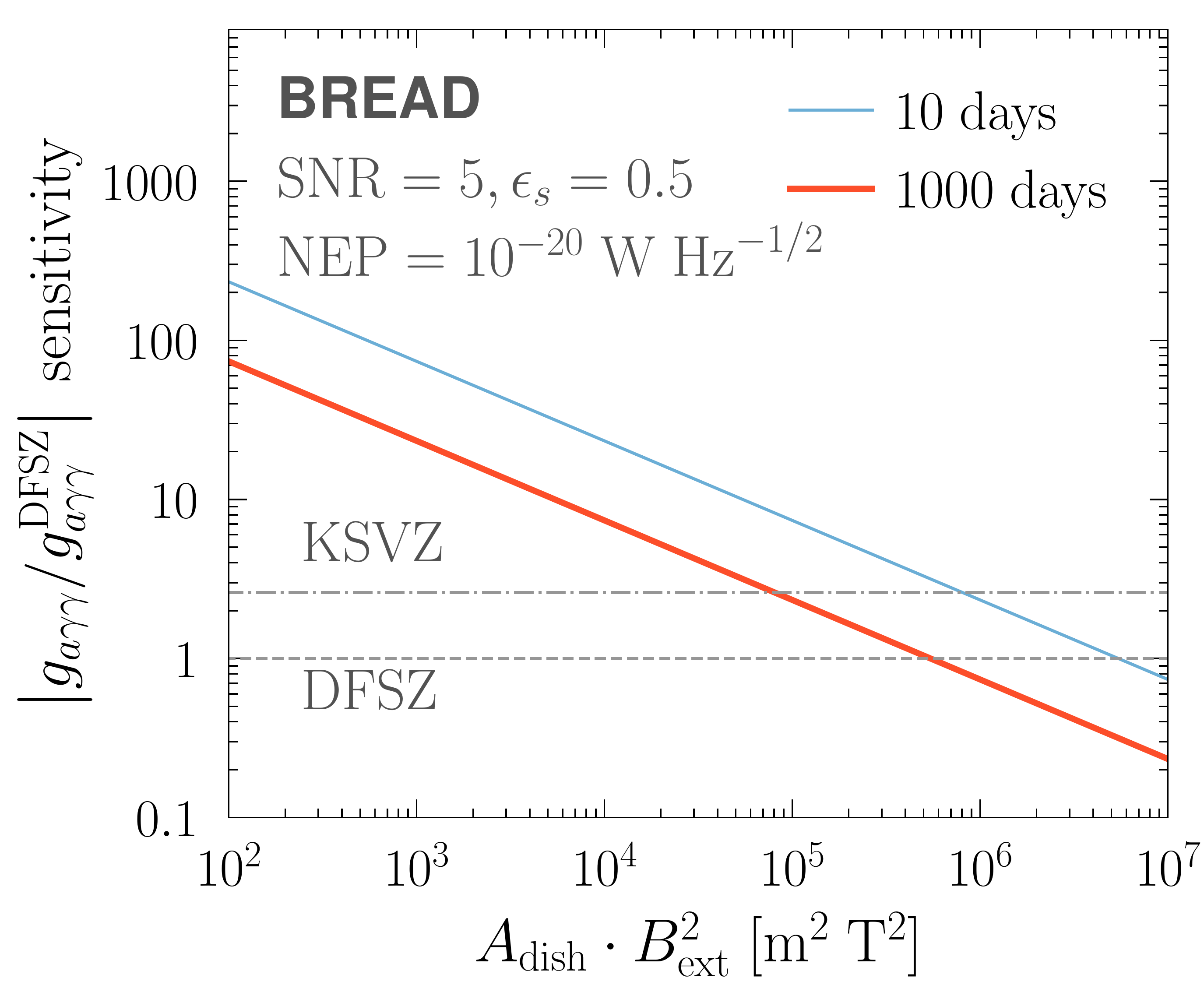}%
    \includegraphics[width=0.33\linewidth]{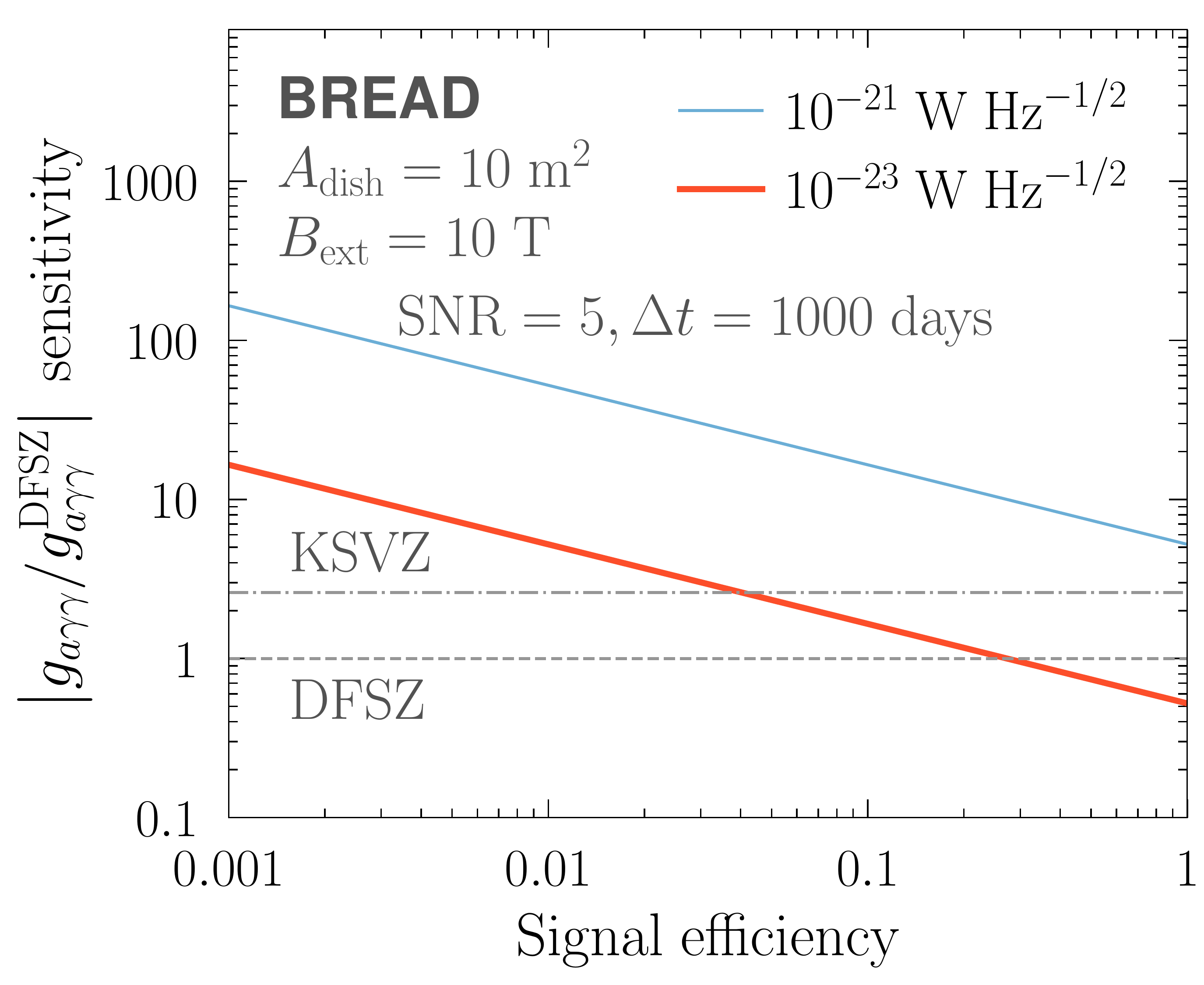}%
    \includegraphics[width=0.33\linewidth]{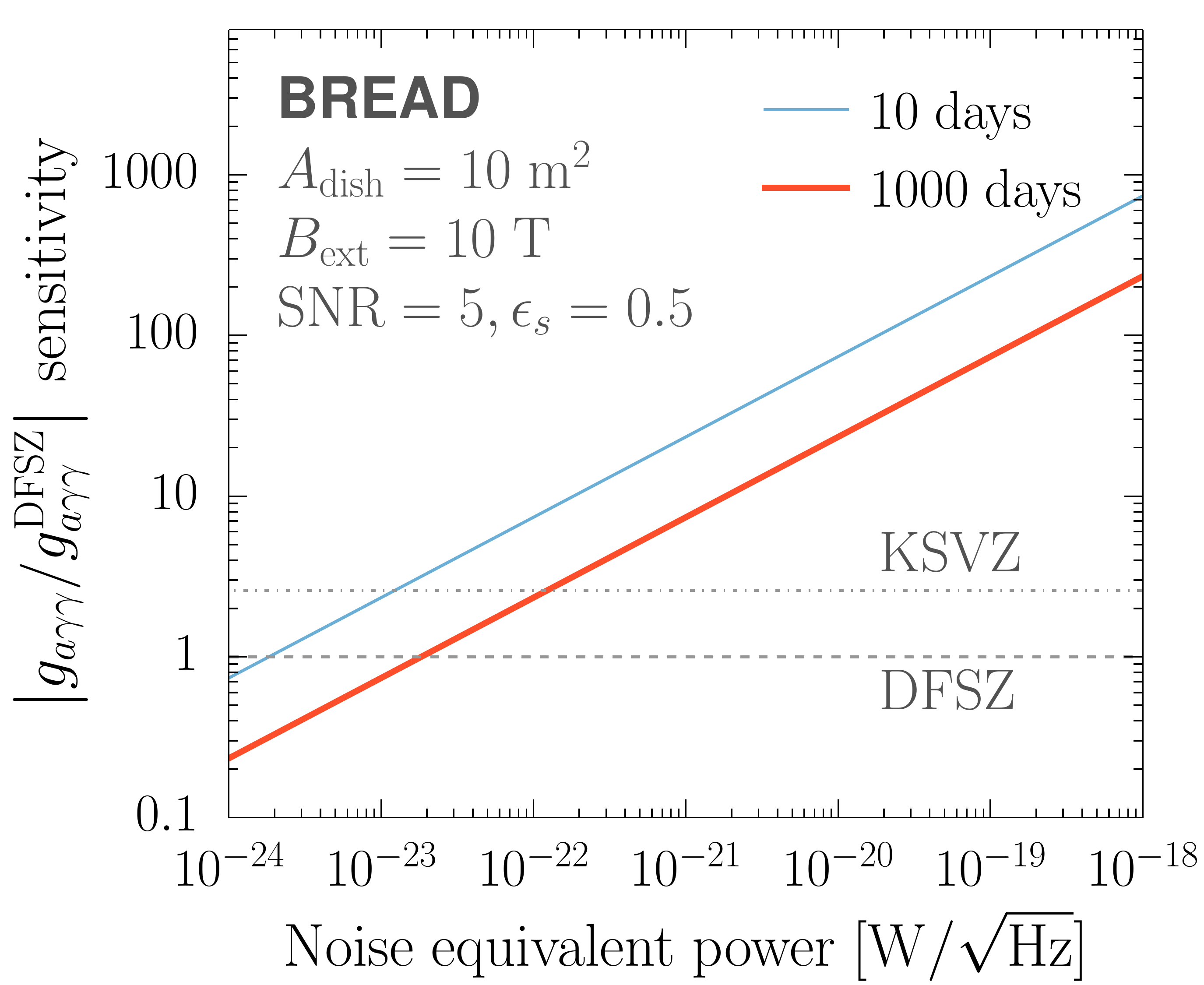}
    \caption{Sensitivity (${\rm SNR = 5}$) to axion--photon coupling normalized to the DFSZ prediction $\left|g_{a\gamma\gamma} / g_{a\gamma\gamma}^\text{DSFZ}\right|$ as a function of various experimental parameters following Eq.~\eqref{eq:axion_dp_sensitivity}: 
    (left) product of dish area and magnetic field $A_\text{dish}\cdot B_\text{ext}^2$, 
    (center) signal emission-to-detection efficiency, 
    (right) noise equivalent power (NEP).
    For the $A_\text{dish}\cdot B_\text{ext}^2$ and NEP plots, two integration times of 10 and 1000 days are shown. 
    For the signal efficiency plot, two NEP values of $10^{-21}$ and $10^{-23}$\,W\,Hz$^{-1/2}$ are considered. 
    The KSVZ (DFSZ) predictions are shown as dot-dashed (dashed) gray horizontal lines. }
    \label{fig:axion_coupling_varyexp}
\end{figure*}

In photon counting regimes, it is more convenient to consider the DM-induced rate $R_\text{DM}$ of emitted photons given by $P_\text{DM}/m_\text{DM}$: 
\begin{linenomath}
\begin{align}
    \left\{\!
    \begin{array}{c}
        \dfrac{R_a   }{0.55\,\text{Hz}}   \\[0.3cm]
        \dfrac{R_{A'}}{0.14\,\text{Hz}}
    \end{array} 
    \!\right\}
    &= \left\{\!\!
    \begin{array}{c}
	    \left(\dfrac{g_{a\gamma\gamma}}{10^{-11}\,\text{GeV}^{-1}}\right)^2
	    \left(\dfrac{\text{meV}}{m_a}\right)^3 
        \left(\dfrac{B_\text{ext}}{10\,\text{T}}\right)^2 \\[0.3cm]
        \dfrac{\alpha_\text{pol}^2}{2/3} \left(\dfrac{\kappa}{10^{-14}}\right)^2
        \dfrac{\text{meV}}{m_{A'}}
    \end{array} 
    \!\!\right\}
    \nonumber\\
    &  \quad \times 
    \frac{\rho_\text{DM}}{0.45\,\text{GeV/cm}^3}
	\frac{A_\text{dish}}{10\,\text{m}^2}.
    \label{eq:axion_dp_rate}
\end{align}
\end{linenomath}

For counting statistics, the relevant figure-of-merit is the significance $Z$ given a runtime $\Delta t$, which for a photosensor with dark count rate DCR, is estimated as
\begin{equation}
    Z = \frac{N_\text{signal}}{\sqrt{N_\text{noise}}} =\frac{\epsilon_s R_\text{DM} \Delta t}{\sqrt{\text{DCR} \Delta t}}.
\end{equation}
Formally, $N_\text{signal}/\sqrt{N_\text{noise}}$ holds in the Gaussian regime where $N_\text{noise} \gtrsim 10$, below which Poissonian statistics applies.
For zero noise counts $N_\text{noise} = 0$, one has 95\% CL exclusion sensitivity for three or more signal events. 

Requiring $Z = 5$ for DM reach implies the coupling sensitivity is related to the DCR by 
\begin{linenomath}
\begin{align}
    \left\{\!\!
    \begin{array}{c}
        \left(\dfrac{g_{a\gamma\gamma}}{10^{-12}}\right)^2   \\
         \left(\dfrac{\kappa}{10^{-15}}\right)^2  
    \end{array} 
    \!\!\right\}
    &= \left\{\!\!
    \begin{array}{c}
        \dfrac{3.0}{\text{GeV}^2}\!
    \left(\dfrac{m_a}{\text{meV}}\right)^3
    \left(\dfrac{10\,\text{T}}{B_\text{ext}}\right)^{2}  \\
       11.9\dfrac{2/3}{\alpha_\text{pol}^2} \dfrac{m_{A'}}{\text{meV}} 
    \end{array} 
    \!\!\right\}\!
    \left(\frac{\text{hour}}{\Delta t}\right)^{1/2}
    \nonumber\\
    & \!\!\!\!\!\!\!\!\! \times \!
    \frac{10\,\text{m}^2}{A_\text{dish}}
    \frac{Z}{5}
    \frac{0.5}{\epsilon_s} 
    \left(\frac{\text{DCR}}{10^{-2}\,\text{Hz}}\right)^{1/2}
    \frac{0.45\,\text{GeV/cm}^3}{\rho_\text{DM}}.
    \label{eq:axion_dp_sensitivity_counting}
\end{align}
\end{linenomath}
While measuring DCR is more relevant for photosensors operating in photon counting regimes, device physics literature often converts to NEP using $\text{NEP} = (E/\epsilon)\sqrt{2\cdot\text{DCR}}$ to facilitate comparisons~\cite{Chen:18}, where $\epsilon$ is the detection efficiency. 

Canonical QCD axion scenarios~\cite{GrillidiCortona:2015jxo} fix the relation between the axion coupling and mass to $g_{a\gamma\gamma}^\text{KSVZ} = -3.9 \times 10^{-13} (m_a/\text{meV})~\text{GeV}^{-1}$ and $g_{a\gamma\gamma}^\text{DFSZ} = 1.5 \times 10^{-13} (m_a/\text{meV}) ~\text{GeV}^{-1}$.
The prefactor depends on the theoretical details of the ultraviolet completion assumed in the KSVZ and DFSZ models~\cite{Dine:1981rt,Zhitnitsky:1980tq,Kim:1979if,Shifman:1979if}. 
Using Eq.~\eqref{eq:axion_dp_rate}, we find the photon emission rate is 
\begin{align}
   \left\{\begin{array}{c}
   R_a^\text{KSVZ}\\[0.3cm]
   R_a^\text{DFSZ}
   \end{array}\right\}
   &=
   \left\{
    \begin{array}{c}
	    \dfrac{72.2}{{\rm day}}
        \\[0.3cm]
	    \dfrac{10.6}{{\rm day}}
    \end{array} 
    \right\}
   \dfrac{\text{meV}}{m_a}
   \left(\dfrac{B_\text{ext}}{10\,\text{T}}\right)^2\nonumber \\
   &\quad \quad \times 
   \frac{\rho_\text{DM}}{0.45\,\text{GeV\,cm}^{-3}}
	\frac{A_\text{dish}}{10\,\text{m}^2}.
	\label{eq:summary_qcd_axion_rate}
\end{align}
Figure~\ref{fig:photon_rate_qcdaxion} shows the photon emission rate $R_\gamma = P_\text{DM} / m_\text{DM}$ for the QCD axion scenarios assuming $A_\text{dish} = 10$\,m$^2$ and $B_\text{ext} = 10$\,T, where the impact of reducing these parameters by a factor of ten is shown. 
In this configuration, we estimate on the order of 100 photons per day for 1\,THz (4\,meV) photons considering the KSVZ scenario. 
Higher masses $m_a \gtrsim 1$\,eV requires longer runtimes $\Delta t \gtrsim {\rm months}$ to overcome shot noise.

\subsection{Cryostat and magnet considerations}
The BREAD geometry is advantageous over recent dish antenna designs that use spherical or flat emitting surfaces~\cite{Suzuki:2015sza,Knirck:2018ojz,Tomita:2020usq,FUNKExperiment:2020ofv,BRASS:website} because the latter are not suited for enclosure in a solenoid and typically require custom magnet designs.
Solenoids are the leading magnet design that allows access to large bore sizes and high magnetic fields. 
An extant large-bore of radius $R = 3$\,m solenoid in particle physics is the CMS experiment at CERN but with modest fields $B = 3.8$\,T, while the ITER project features an $R=1.3$\,m, $B = 13$\,T magnet~\cite{BUDINGER2018509,Bird:2020nos}. 
However, these meter-scale magnets cost around three orders of magnitude more than magnets used for state-of-the-art DM experiments e.g.\ ADMX. 

The dish area and magnetic field are two haloscope parameters that bound the instrument dimensions and govern experimental sensitivity of BREAD. 
The coupling sensitivity improves as $\sim A_\text{dish}^{1/2}B_\text{ext}$. 
One may consider advances in both existing cryostat and magnets size possible with 10 years of research-and-development.
Figure~\ref{fig:axion_coupling_varyexp} (left) shows how the sensitivity to the axion--photon coupling varies with the product $A_\text{dish} \cdot B_\text{ext}^2$ for 1 and 1000 days runtime.
For the proposed aspect ratio where the barrel length $L$ is related to its radius $R$ by $L=2\sqrt{2}R$,
the radius is fully determined from the dish area $A_\text{dish}$ by $R = \sqrt{A_\text{dish}/(4\pi \sqrt{2})}$.
So as $R \sim A^{1/2}$ for BREAD, coupling sensitivity scales linearly with both bore radius and field strength $B_\text{ext} R$.  

The radius for the pilot $A_\text{dish} = 0.7$\,m$^2$ is $R = 0.2$\,m and the nominal experiment with $A_\text{dish} = 10$\,m$^2$ has $R = 0.75$\,m.
While a $A_\text{dish} = 100$\,m$^2$ barrel that has $R = 2.4$\,m is within engineering feasibility, the financial cost of the large cryostat and magnet required is challenging.
The energy stored in the magnetic field in the solenoid is $(B^2V)/(2\mu_0) = (\pi 2\sqrt{2} R^3B^2 )/(2\mu_0)$, where $\mu_0$ is the vacuum magnetic permeability, which corresponds to 150\,MJ for $R=0.75\,$m, $B=10\,$T.
For comparison, standard medical magnetic-resonance-imaging (MRI) magnets typically have a bore radius of around 0.35\,m.
Constructing a dedicated solenoid for BREAD is financially impractical. 
We have identified an available 9.4\,T solenoid originally constructed for MRI research at the University of Illinois at Chicago, which we plan to repurpose for next-generation axion DM experiments at Fermilab including BREAD. 
Given coupling sensitivity scales quickest with the magnetic field, this reinforces the motivation for continued research-and-development into high-field large-bore solenoids. 
Once the haloscope design parameters $A_\text{dish}, B_\text{ext}$ are maximized within engineering and financial feasibility, photosensor performance then governs DM discovery reach.

\begin{figure*}[tbh]
    \centering
    \includegraphics[width=0.5\linewidth]{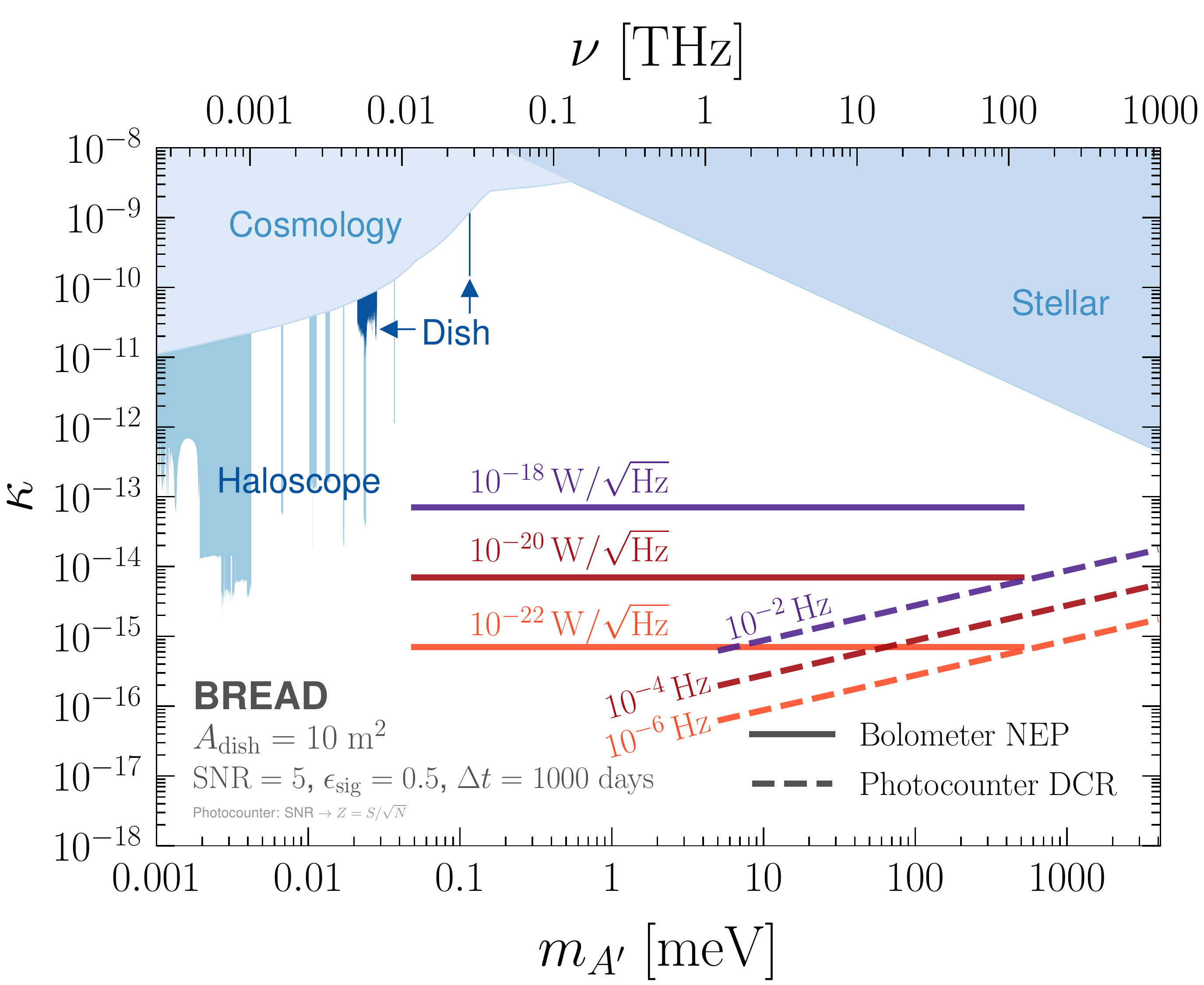}%
    \includegraphics[width=0.5\linewidth]{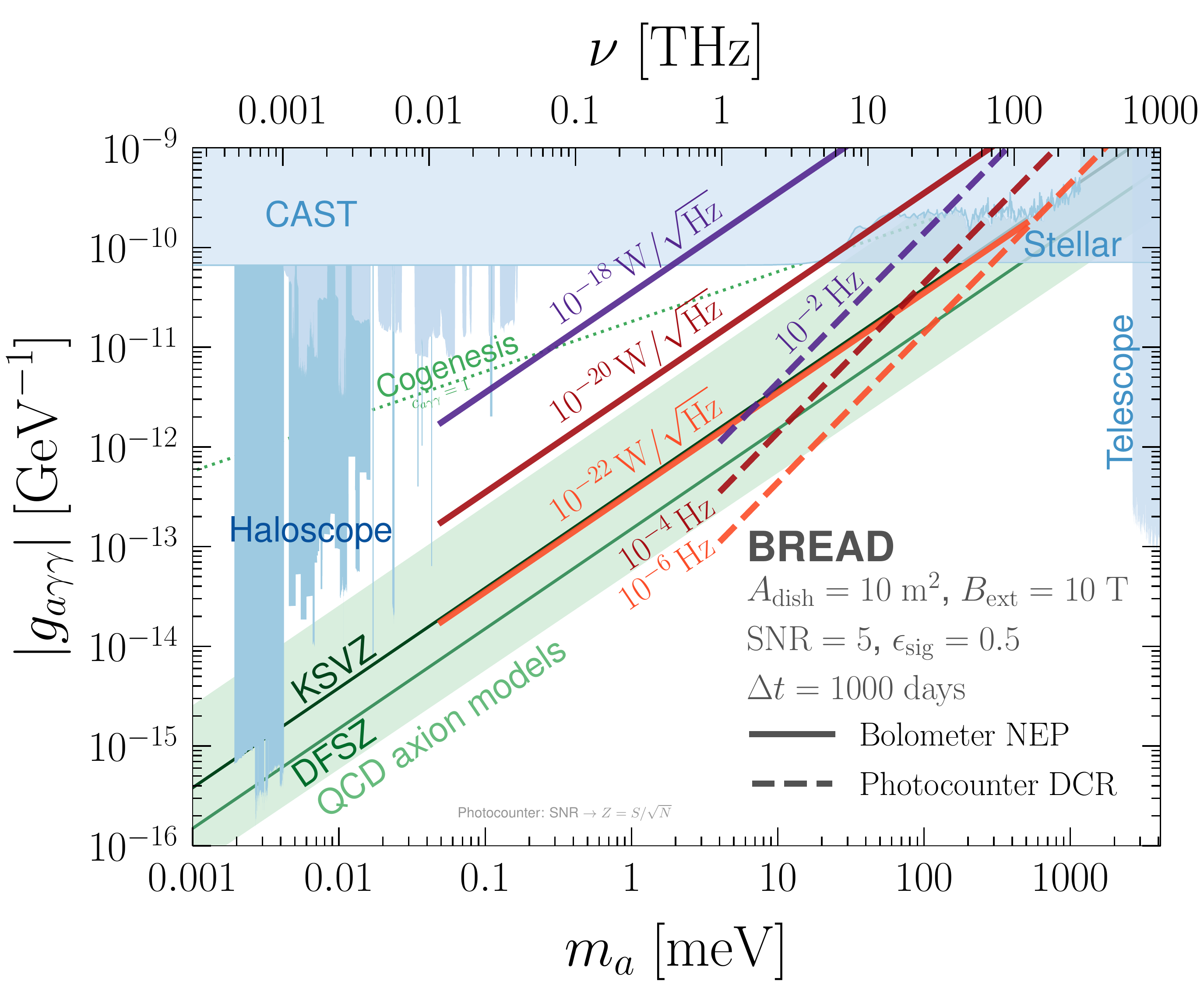}
    \caption{
    Projected sensitivity for dark photons $A'$ (left) and axions $a$ (right) assuming generic bolometers for different noise equivalent power (NEP) and photocounters for various dark count rate (DCR) with $\Delta t = 1000$\,days runtime. 
    Analogous to Fig.~\ref{fig:sensitivity_photosensors}, 
    blue shading shows existing constraints from Ref.~\cite{Caputo:2021eaa} while benchmark axion predictions include QCD axion models~\cite{DiLuzio:2016sbl} (green band), KSVZ~\cite{Kim:1979if,Shifman:1979if} (dark green line), DFSZ~\cite{Dine:1981rt,Zhitnitsky:1980tq} (lighter green line), and cogenesis~\cite{Co:2020xlh} (green dots). 
    }
    \label{fig:axion_general_varyNEP}
\end{figure*}

\subsection{Focusing and velocity effects\label{apndx:velocity_focusing}}

In principle, the parabolic reflector focuses EM radiation emitted perpendicularly to the barrel onto a point, but in practice, the resulting spot size is smeared. 
This is due to a nonzero DM velocity causing the direction of photon emission to deviate from the surface normal by a small angle. 
Therefore, if the photosensor area $A_\text{sens}$ is smaller than the signal spot area $A_\text{spot}$, $A_\text{sens} < A_\text{spot}$, the signal detection efficiency is reduced. 

\begin{figure}[tbh]
    \centering
    \includegraphics[width=\linewidth]{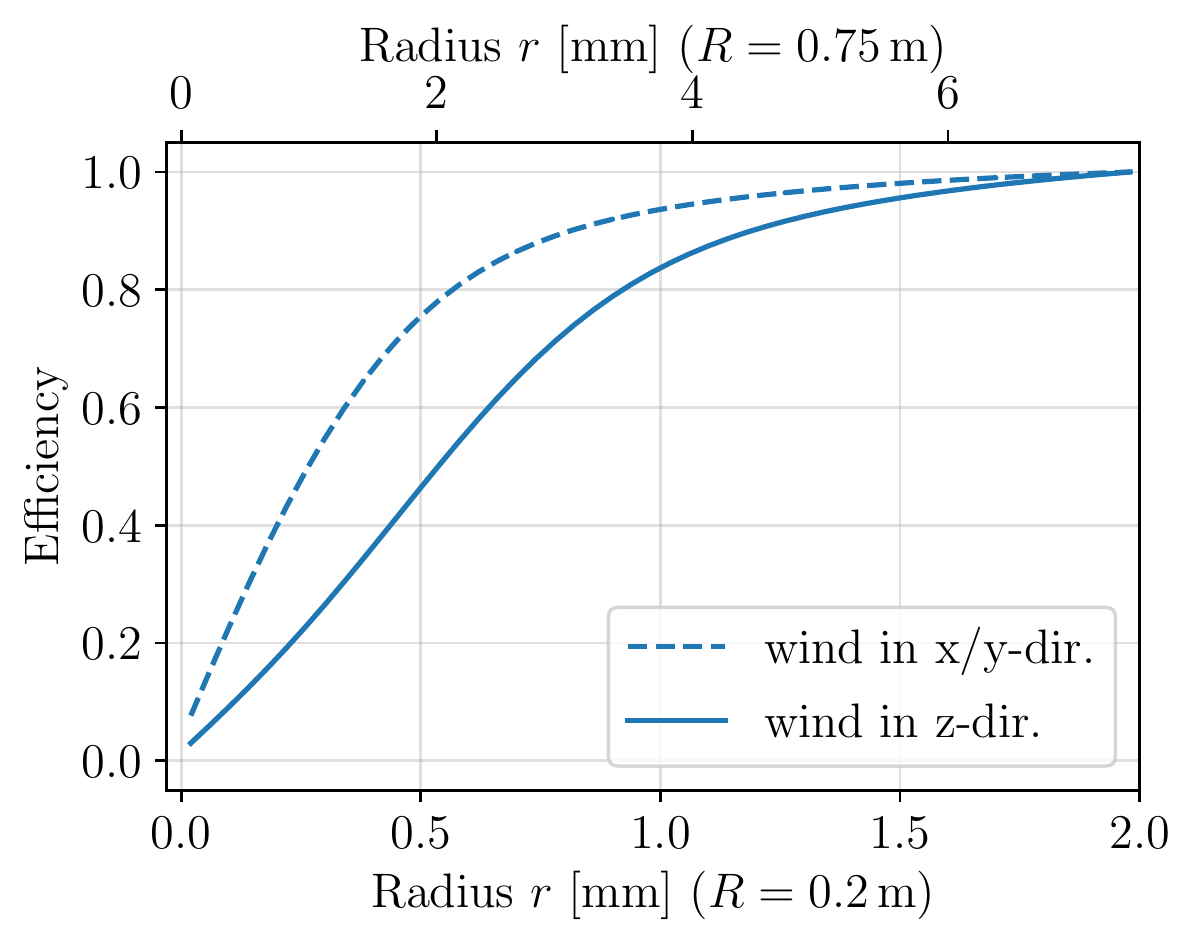}
    \caption{Detection efficiency in the short-wavelength limit as a function of detector radius $r$ due to DM velocity effects smearing out the focal spot. 
    The solid (dashed) blue line corresponds to the efficiency assuming the DM wind velocity is aligned with the $z$ ($x/y$) axes.
    }
    \label{fig:velocity_spotsize_efficiency}
\end{figure}

Figure~\ref{fig:velocity_spotsize_efficiency} shows the impact of DM velocity on the geometric signal efficiency $A_\text{sens}/A_\text{spot}$ assuming the sensor has saturated efficiency i.e.\ each photon that arrives within $A_\text{sens}$ is absorbed and a detection reported.
The upper (lower) axis shows the detector radius assuming the barrel radius is $R=0.75~(0.2)$\,m. 
The dashed (solid) line shows the most optimistic (pessimistic) scenario for the DM wind velocity aligned in the $z$ ($x/y$) direction.
For the pilot experiment with $R=0.2$\,m with $A_\text{dish} = 0.7$\,m$^2$, achieving 50\% signal efficiency assuming $z$ ($x/y$) DM-wind alignment requires a 0.25\,mm (0.5\,mm) sensor radius.
The larger nominal radius of $R=0.75$\,m requires a 1\,mm (2\,mm) radius sensor.
As discussed in the main text, such sensor dimensions are readily fabricated for commercial bolometers but state-of-the-art superconducting photosensors are typically on the order of $\sim 0.1$\,mm in size or smaller. 
Figure~\ref{fig:axion_coupling_varyexp} (center) shows how the sensitivity varies with detection efficiency, showing the modest $\sqrt{\epsilon_s}$ dependence in Eq.~\eqref{eq:axion_dp_sensitivity}. 
Indeed, the main text sets an efficiency of $\epsilon_s = 50\%$ for simplicity. 
Because the coupling sensitivity from Eq.~\eqref{eq:axion_dp_sensitivity_counting} only scales as $\sqrt{\epsilon_s}$, if instead $\epsilon_s = 5\%$, the sensitivity is only reduced by a factor of around three.
Nonetheless, this sets an important BREAD physics-driven instrumentation goal of scaling the candidate sensors to millimeter dimensions while maintaining low noise. 

\subsection{Photosensor performance and noise}

The main text identifies various candidate technologies that could meet the stringent requirements of BREAD. 
The values reported in the literature are often optimized to certain spectral bands or optical configurations, and while the sensor technologies typically have broadband response, this requires experimental demonstration.
Photosensors are also typically fabricated as a flat plane for perpendicular illumination, whereas the photons in BREAD can arrive at steep incidence angles. 
As the photons are emitted perpendicularly to the surface (and magnetic field), they also have specific polarization. 
Specifically for BREAD, it is important to characterize how both signal efficiency and noise (dark count) performance scales with not only frequency but also angle-of-incidence and polarization. 
Many of these photosensor properties are challenging to simulate, therefore dedicated in-situ measurements of candidate photosensor technologies are required and under preparation. 
Ensuring signal efficiency and noise performance are demonstrated while scaling sensor active area to millimeter-sized pixels and/or arrays are key research goals for BREAD.

\begin{table}[tbh]
\centering
	\begin{tabular}{lcccc}
	\toprule
	\textbf{BREAD}          & Pilot       & Stage 1     & Stage 2a     & Stage 2b   \\
    \midrule
	Axion $a$               & ---         &$\checkmark$ & $\checkmark$ & $\checkmark$\\
	Dark photon $A'$        & $\checkmark$&$\checkmark$ & $\checkmark$ & $\checkmark$\\
	\midrule
    \multicolumn{4}{l}{Experimental parameters} \\
    \midrule
    $A_\text{dish}$ [m$^2$] & 0.7         & 10          & 10           & 10 \\
	$B_\text{ext}$ [T]      & ---         & 10          & 10           & 10 \\
	$\epsilon_s$            & 0.5         & 0.5         & 0.5          & 0.5 \\
	$\Delta t$ [days]       & 10          & 10          & 1000         & 1000\\
	NEP [W\,Hz$^{-1/2}$]    & $10^{-14}$  & $10^{-18}$  & $10^{-20}$   & $10^{-22}$\\
	\midrule
	\multicolumn{4}{l}{Coupling sensitivity (SNR = 5)} \\
	\midrule
	$\left|g_{a\gamma\gamma} / g_{a\gamma\gamma}^\text{KSVZ}\right|$ & ---  & 280 & 9.0 & 0.90 \\
	$\left|g_{a\gamma\gamma} / g_{a\gamma\gamma}^\text{DFSZ}\right|$ & ---  & 740 & 23  & 2.3  \\
	$\kappa/10^{-14}$                                                & 8400 & 22  & 0.7 & 0.07 \\
	\bottomrule
	\end{tabular}
	\caption{\label{tab:exp_parameters}Benchmark experimental parameters for staging the BREAD science program. 
	Shown are the dish area $A_\text{dish}$, external magnetic field $B_\text{ext}$, signal emission-to-detector efficiency $\epsilon_s$, runtime $\Delta t$, and photosensor noise equivalent power (NEP). 
	The lower rows show the sensitivity assuming signal-to-noise ratio SNR $=5$ to axion--photon couplings $g_{a\gamma\gamma}$ normalized to the KSVZ/DFSZ predictions and dark photon kinetic mixing $\kappa$ using Eq.~\eqref{eq:axion_dp_sensitivity}.  }
\end{table}

The noise budget and sources can be categorized by:  
\emph{intrinsic} to sensor such as readout, internal thermal fluctuations; 
\emph{extrinsic} to sensor such as environmental thermal emission and cosmic rays.
Devices that act like bolometers are limited by intrinsic sources such as thermal fluctuations in TESs, and generation-recombination noise of pairs in KIDs. 
Single photon counters such as SNSPDs are reaching very low noise that they are limited by extrinsic sources. 
Carefully understanding these noise sources experimentally for BREAD could help design strategies for their mitigation. 

Environmental conditions that would affect long-term low-background measurements are challenges not unique to BREAD but also other DM experiments. 
Cosmic-ray backgrounds are stochastic, are not expected to have long-term variations, and can be actively rejected with in situ veto systems. 
The alignment of the barrel-reflector-sensor setup could be subject to time-dependent environmental variations such as temperature and humidity or mechanical vibrations.
These can be rejected by correlation with active monitoring systems, and further accommodated with regular dedicated calibration and alignment runs. 
Detailed studies of this requires in situ data of the experimental site that is the subject of future work.

To provide concrete targets driven by BREAD for photosensor noise, we can estimate the required NEP for sensitivity to the KSVZ and DFSZ axion assuming different runtimes in 
Figure~\ref{fig:axion_coupling_varyexp} (right) following Eq.~\eqref{eq:axion_dp_sensitivity}.
The state-of-the-art NEP today is around $3 \times 10^{-21}{\rm \,W\,Hz}^{-1/2}$ for quantum capacitance detectors~\cite{Echternach:2018}. 
There are also recent claims TES-based device could achieve $8 \times 10^{-22}$\,W\,Hz$^{-1/2}$ electrical NEP~\cite{Nagler:2020mki}. 
With this noise performance, it remains challenging to probe the KSVZ scenario assuming 1000 days runtime, which requires $1 \times 10^{-22}{\rm \,W\,Hz}^{-1/2}$. 
Probing DFSZ requires an order of magnitude improvement in NEP to $2 \times 10^{-23}{\rm \,W\,Hz}^{-1/2}$, which is very challenging but motivates a target for sensor research-and-development.  

Figure~\ref{fig:axion_general_varyNEP} shows the projected reach for dark photons and axions in the coupling vs.\ mass plane for generic bolometric photosensors or photocounters, assuming various NEP and DCR for $\Delta t = 1000$\,days runtime. 
Experimentally demonstrating that the required NEP is achievable across several decades of frequency with high efficiency is a challenging requirement for BREAD to probe QCD axion models.
The scientific potential for realizing this is significant, given the multidecade mass sensitivity without needing to tune the experiment to an unknown DM mass. 
Photocounters such as SNSPDs have promising dark count rates, but their low-mass (longer wavelength) reach are currently limited to $E > 125$\,meV photon energies ($\lambda < 10$\,$\mu$m wavelengths)~\cite{Verma:2020gso}. 
Lowering these photon energy thresholds is a key sensor research-and-development target that would have a significant payoff in BREAD.

\subsection{Further bolometry discussion}

We expand on details related to bolometers.
For the SNR calculation with bolometers, the bandwidth of the sensor does not enter directly. 
A chopper or moving mirror modulates the signal, allowing noise mitigation by rejecting DC power components.
Thus the bandwidth appearing in the signal-to-noise calculation is the bandwidth of the modulated signal, which becomes narrower with integration time, giving an SNR that improves with $\sqrt{\Delta t}$. 
The sensor bandwidth of a bolometer is measured with pulse rise and decay time.
This is only relevant in that it restricts the speed that the signal can be modulated with a chopper. 
For example, the chopper acquired with the \textsc{IR Labs} bolometer operates at 100\,Hz implying a 10\,millisecond or faster bolometer response is required. 

The noise equivalent power NEP is related to the thermal relaxation time of a bolometer.
Therefore, the design goals of NEP and large bandwidth are in tension and the lowest noise bolometers will be slow. 
Bolometers need to have sufficiently fast response such that a chopper can modulate the signal at a frequency that makes $1/f$ components of the system noise subdominant.
We anticipate this does depend on the specific device technology, and characterizing this for candidate sensors with experimental measurements is an important part of the BREAD research program.

\subsection{Experimental staging}

Table~\ref{tab:exp_parameters} summarizes benchmark experimental parameters for a staged approach to BREAD. 
The corresponding sensitivity to the axion $g_{a\gamma\gamma}$ and dark photon $\kappa$ couplings are normalized to benchmark targets. 
We consider the prototypical pilot dark photon search with 10 days runtime.
This demonstrator forms a nearer-term science goal that will give valuable experimental experience, already provide meaningful physics results,
and serve as a proof-of-principle for BREAD.
Longer term, stage 1 shows the nominal experiment with the full-scale experiment inside a 10\,T magnet assuming only a $10^{-18}$\,W\,Hz$^{-1/2}$ NEP photosensor can be installed and demonstrated, which can start to probe currently unconstrained axion parameter space. 
The planned stage 2a (2b) considers 1000 days runtime and experimental work to successfully develop and couple a $10^{-20}~(10^{-22})$\,W\,Hz$^{-1/2}$ NEP photosensor.
This could probe QCD axion scenarios assuming ongoing research-and-development can demonstrate multiple order-of-magnitude improvements in sensor NEP.

\end{document}